\documentclass{article}
\usepackage{url} 
\usepackage{enumerate, stmaryrd,mathrsfs}
\usepackage{amsmath,amssymb,latexsym}
\usepackage{times} 

\newcounter{zaehler_enumerate_kurz}

\newcommand*{\twodots}{.\,.\,}
\newcommand*{\twocdots}{{\cdot}\,{\cdot}\,}

\newcommand*{\raus}[1]{ }

\newcommand*{\parno}{\par\noindent}

\newcommand*{\vek}[1]{\ensuremath{\vec{#1}}}

\newcommand*{\bs}[1]{\ensuremath{\boldsymbol{#1}}}

\newcommand*{\und}{\ensuremath{\wedge}}
\newcommand*{\Und}{\ensuremath{\bigwedge}}
\newcommand*{\oder}{\ensuremath{\vee}}
\newcommand*{\Oder}{\ensuremath{\bigvee}}
\newcommand*{\nicht}{\ensuremath{\neg}}
\newcommand*{\impl}{\ensuremath{\rightarrow}}
\newcommand*{\gdw}{\ensuremath{\leftrightarrow}}

\renewcommand*{\geq}{\ensuremath{\geqslant}}
\renewcommand*{\leq}{\ensuremath{\leqslant}}

\newcommand*{\set}[1]{\ensuremath{\{ #1 \}}}
\newcommand*{\setc}[2]{\set{#1 \,:\, #2}}

\newcommand*{\bigset}[1]{\ensuremath{ \left\{ #1 \right\} } }
\newcommand*{\bigsetc}[2]{\bigset{#1 \,:\, #2}}

\newcommand*{\dom}[1]{\ensuremath{\underline{#1}}}

\newcommand*{\FktsFont}[1]{\ensuremath{\textup{#1}}}

\newcommand*{\lcm}{\ensuremath{\FktsFont{lcm}}}

\newcommand*{\abgerundet}[1]{\ensuremath{
    \left\lfloor {#1} \right\rfloor }}
\newcommand*{\aufgerundet}[1]{\ensuremath{\left\lceil {#1} \right\rceil }}

\newcommand*{\deff}{:=}

\newcommand*{\NN}{\ensuremath{\mathbb{N}}}    
\newcommand*{\NNpos}{\ensuremath{\NN_{\mbox{\tiny $\scriptscriptstyle > 0$}}}}

\newcommand*{\ZZ}{\ensuremath{\mathbb{Z}}}

\newcommand*{\RR}{\ensuremath{\mathbb{R}}}
\newcommand*{\QQ}{\ensuremath{\mathbb{Q}}}

\newcommand*{\UU}{\ensuremath{\mathbb{U}}}

\newcommand*{\A}{\ensuremath{\mathcal{A}}}
\newcommand*{\B}{\ensuremath{\mathcal{B}}}

\newcommand*{\K}{\ensuremath{\mathscr{K}}}

\newcommand*{\letter}[1]{\ensuremath{\texttt{\upshape #1}}} 

\newcommand*{\bigO}{\ensuremath{{O}}}

\newfont{\MyScript}{cmfi10 at 10pt}

\newcommand*{\ar}{\ensuremath{\textit{ar}}}

\newcommand*{\Problem}[1]{\ensuremath{\textsc{#1}}}

\newcommand*{\Spectrum}{\Problem{Spec}}

\newcommand*{\Parity}{\Problem{Parity}}

\newcommand*{\Reach}{\Problem{Reach}}
\newcommand*{\detReach}{\Problem{Det-Reach}}
\newcommand*{\Conn}{\Problem{Conn}}

\newcommand*{\struc}[1]{\ensuremath{\langle #1 \rangle}}

\newcommand*{\praedikat}[1]{\ensuremath{\textsl{#1}}}
\newcommand*{\Bit}{\praedikat{Bit}}
\newcommand*{\Squares}{\praedikat{Squares}}

\newcommand*{\Exp}{\praedikat{Exp}}

\newcommand*{\Max}{\praedikat{max}}

\newcommand*{\Praedikatklasse}[1]{\ensuremath{\mathfrak{#1}}}

\newcommand*{\Arb}{\Praedikatklasse{Arb}}
\newcommand*{\Bip}{\Praedikatklasse{Bip}} 
\newcommand*{\Num}{\Bip} 

\newcommand*{\Presb}{\Praedikatklasse{Presb}}

\newcommand*{\Nr}{\#}
\newcommand*{\Count}{\#}

\newcommand*{\free}{\ensuremath{\textit{free\,}}}

\newcommand*{\class}[1]{\ensuremath{\textsl{#1}}}

\newcommand*{\FO}{\class{FO}}
\newcommand*{\FOunC}{\class{FOunC}}
\newcommand*{\FObinC}{\class{FObinC}}
\newcommand*{\FOkaryC}{\class{FO}k\class{-aryC}}
\newcommand*{\FOonearyC}{\class{FO}1\class{-aryC}}
\newcommand*{\FOtwoaryC}{\class{FO}2\class{-aryC}}

\newcommand*{\FOunM}{\class{FOunM}}
\newcommand*{\FObinM}{\class{FObinM}}

\newcommand*{\LFP}{\class{LFP}}

\newcommand*{\MSO}{\class{MSO}}

\newcommand*{\NP}{\class{NP}}

\newcommand*{\ETIME}{\class{ETIME}}
\newcommand*{\LinH}{\class{LINH}}
\newcommand*{\Logspace}{\class{LOGSPACE}}

\newcommand*{\TCzero}{\ensuremath{\class{TC}^0}}
\newcommand*{\ACzero}{\ensuremath{\class{AC}^0}}

\usepackage{amsthm} 

\swapnumbers  

\newcommand*{\fertig}{{}\hfill \ensuremath{\square}}
\renewcommand*{\qed} {\null\hfill\ensuremath{\blacksquare}}

\newtheorem{quasi_theorem}{Theorem}[section] 
\newtheorem*{quasi_theorem_nn}{Theorem} 

\newtheorem{quasi_lemma}[quasi_theorem]{Lemma}
\newtheorem*{quasi_lemma_nn}{Lemma} 

\newtheorem{quasi_corollary}[quasi_theorem]{Corollary}
\newtheorem*{quasi_corollary_nn}{Corollary} 

\newtheorem{quasi_proposition}[quasi_theorem]{Proposition}
\newtheorem*{quasi_proposition_nn}{Proposition} 

\newtheorem{quasi_maintheorem}[quasi_theorem]{Theorem}
\newtheorem*{quasi_maintheorem_nn}{Theorem} 

\newtheorem{quasi_definition}[quasi_theorem]{Definition}
\newtheorem*{quasi_definition_nn}{Definition} 

\newtheorem{quasi_example}[quasi_theorem]{Example}
\newtheorem*{quasi_example_nn}{Example} 

\newtheorem{quasi_remark}[quasi_theorem]{Remark}
\newtheorem*{quasi_remark_nn}{Remark} 

\newtheorem{quasi_remarks}[quasi_theorem]{Remarks}
\newtheorem*{quasi_remarks_nn}{Remarks} 

\newtheorem{quasi_question}[quasi_theorem]{Question}
\newtheorem*{quasi_question_nn}{Question} 

\newtheorem{quasi_fact}[quasi_theorem]{Fact}
\newtheorem*{quasi_fact_nn}{Fact} 

\newtheorem{quasi_facts}[quasi_theorem]{Facts}
\newtheorem*{quasi_facts_nn}{Facts} 

\newenvironment{theorem_ohne}{\begin{quasi_theorem}}{\end{quasi_theorem}\noindent}
\newenvironment{theorem_nn_ohne}{\begin{quasi_theorem_nn}}{\end{quasi_theorem_nn}\noindent}

\newenvironment{lemma_ohne}{\begin{quasi_lemma}}{\end{quasi_lemma}\noindent}
\newenvironment{lemma_nn_ohne}{\begin{quasi_lemma_nn}}{\end{quasi_lemma_nn}\noindent}

\newenvironment{corollary_ohne}{\begin{quasi_corollary}}{\end{quasi_corollary}\noindent}
\newenvironment{corollary_nn_ohne}{\begin{quasi_corollary_nn}}{\end{quasi_corollary_nn}\noindent}

\newenvironment{proposition_ohne}{\begin{quasi_proposition}}{\end{quasi_proposition}\noindent}
\newenvironment{proposition_nn_ohne}{\begin{quasi_proposition_nn}}{\end{quasi_proposition_nn}\noindent}

\newenvironment{maintheorem_ohne}{\begin{quasi_maintheorem}}{\end{quasi_maintheorem}\noindent}
\newenvironment{maintheorem_nn_ohne}{\begin{quasi_maintheorem_nn}}{\end{quasi_maintheorem_nn}\noindent}

\newenvironment{definition_ohne}{\begin{quasi_definition}}{\end{quasi_definition}\noindent}
\newenvironment{definition_nn_ohne}{\begin{quasi_definition_nn}}{\end{quasi_definition_nn}\noindent}

\newenvironment{example_ohne}{\begin{quasi_example}}{\end{quasi_example}\noindent}
\newenvironment{example_nn_ohne}{\begin{quasi_example_nn}}{\end{quasi_example_nn}\noindent}

\newenvironment{remark_ohne}{\begin{quasi_remark}}{\end{quasi_remark}\noindent}
\newenvironment{remark_nn_ohne}{\begin{quasi_remark_nn}}{\end{quasi_remark_nn}\noindent}

\newenvironment{remarks_ohne}{\begin{quasi_remarks}}{\end{quasi_remarks}\noindent}
\newenvironment{remarks_nn_ohne}{\begin{quasi_remarks_nn}}{\end{quasi_remarks_nn}\noindent}

\newenvironment{question_ohne}{\begin{quasi_question}}{\end{quasi_question}\noindent}
\newenvironment{question_nn_ohne}{\begin{quasi_question_nn}}{\end{quasi_question_nn}\noindent}

\newenvironment{fact_ohne}{\begin{quasi_fact}}{\end{quasi_fact}\noindent}
\newenvironment{fact_nn_ohne}{\begin{quasi_fact_nn}}{\end{quasi_fact_nn}\noindent}

\newenvironment{facts_ohne}{\begin{quasi_facts}}{\end{quasi_facts}\noindent}
\newenvironment{facts_nn_ohne}{\begin{quasi_facts_nn}}{\end{quasi_facts_nn}\noindent}

\newenvironment{proof_ohne}{{\rm \bf Proof.}}{}
\newenvironment{proofc_ohne}[1]{{\rm \bf Proof {#1}.}}{}
\newenvironment{proofsketch_ohne}{{\rm \bf Proof (sketch).}}{}

\newenvironment{theorem_mit}{\begin{theorem_ohne}}{\fertig \end{theorem_ohne}}

\newenvironment{lemma_mit}{\begin{lemma_ohne}}{\fertig \end{lemma_ohne}}

\newenvironment{corollary_mit}{\begin{corollary_ohne}}{\fertig \end{corollary_ohne}}

\newenvironment{proposition_mit}{\begin{proposition_ohne}}{\fertig \end{proposition_ohne}}

\newenvironment{definition_mit}{\begin{definition_ohne}}{\fertig \end{definition_ohne}}

\newenvironment{example_mit}{\begin{example_ohne}}{\fertig \end{example_ohne}}

\newenvironment{question_mit}{\begin{question_ohne}}{\fertig \end{question_ohne}}

\newenvironment{fact_mit}{\begin{fact_ohne}}{\fertig \end{fact_ohne}}

\newenvironment{proof_mit}{\begin{proof_ohne}}{\qed \end{proof_ohne}}
\newenvironment{proofc_mit}[1]{\begin{proofc_ohne}{#1}}{\qed \end{proofc_ohne}}

\usepackage{pstricks,pst-node,pstcol}
\usepackage[dvips]{graphicx}         

\newpsobject{showgrid}{psgrid}{subgriddiv=1,griddots=10,gridlabels=6pt}

\definecolor{rot}{named}{Red}%
\definecolor{hellrot}{named}{Red}
\definecolor{gelb}{named}{Goldenrod}%
\definecolor{hellgrau}{named}{Gray}%

\begin{document}

\title{\bf Arithmetic, First-Order Logic, and \\ Counting Quantifiers}
\author{Nicole Schweikardt\thanks{This research was performed while the 
author was employed at the Johannes Gutenberg-Universit\"at Mainz, Germany.} \\
Laboratory for Foundations of Computer Science\\
University of Edinburgh, Scotland, U.K.\\
Email: \url{nisch@informatik.uni-mainz.de} }

\maketitle 
\begin{abstract}
This paper gives a thorough overview of what is known about first-order
logic with counting quantifiers and with arithmetic predicates.
As a main theorem we show that Presburger arithmetic is
closed under unary counting quantifiers. Precisely, this means that for every first-order
formula $\varphi(y,\vek{z})$ over the signature $\set{<,+}$ there is a first-order formula
$\psi(x,\vek{z})$ which expresses over the structure $\struc{\NN,<,+}$ (respectively, over
initial segments of this structure) that the variable $x$ is interpreted exactly by the
number of possible interpretations of the variable $y$ for which the formula 
$\varphi(y,\vek{z})$ is satisfied.
Applying this theorem, we obtain an easy proof of Ruhl's result that reachability 
(and similarly, connectivity) in finite graphs is not expressible in first-order logic with 
unary counting quantifiers and addition. Furthermore, the above result on Presburger
arithmetic helps to show the failure of a particular version of the Crane Beach conjecture.
\end{abstract}

\tableofcontents


%


\section{Introduction}\label{section:Introduction}
In computational complexity theory the complexity of a problem is  measured by the amount
of time or space resources that are necessary for solving a problem on an (idealized)
computational device such as a Turing machine. Fagin's seminal work tied this computational
complexity to the descriptive complexity, i.e., to the complexity (or, the richness) of a
logic that is capable of descirbing the problem. Until now most computational complexity
classes have been characterized in such a descriptive way by logics that are certain
extensions of first-order logic (cf., the textbooks \cite{Immerman,EbbinghausFlum}).
One thing that most of these logics have in common is that they are powerful enough to 
express arithmetic predicates such as $+$, $\times$, or $\Bit$.
\\ 
In \cite{BIS} it was shown that, on finite ordered structures, first-order logic with varying 
arithmetic predicates
corresponds to the circuit complexity class $\ACzero$ with varying uniformity conditions.
However, there are computationally easy problems such as the $\Parity$-problem (asking
whether the number of 1's in the input string is even), that do not belong to $\ACzero$, i.e.,
that are not definable in first-order logic with arbitrary arithmetic predicates. 
In fact, an important feature that first-order logic lacks is the ability to \emph{count}.
\par
Various different ways of
enriching first-order logic with the ability to count have been
examined in the literature.
A usual approach (cf., \cite{Immerman,Etessami,BenediktKeisler}) is
to consider two-sorted structures that consist of a so-called ``vertex
domain'' for the actual structure and an additional ``number
domain'' for the counting results (usually of the same cardinality as the vertex domain) 
which may or may not be equipped with
arithmetic predicates. However, if the actual structure is itself equipped with a
linear ordering, the additional number domain does not give any additional expressivity
(since the number $i$ can be identified with the $i$-th largest element in the vertex domain;
and the arithmetic predicates on the number domain can be translated into the corresponding
predicates on the vertex domain and vice versa). 
In the present paper we will therefore avoid two-sorted structures. Instead, we will use the 
following approach, restricting attention to structures whose universe is either the set
$\NN$ of natural numbers or an initial segment of $\NN$. We enrich first-order logic
by counting quantifiers of the form $\exists^{=x}y$. For an interpretation $a$ of the 
variable $x$, the formula $\exists^{=x}y\,\varphi(y)$ expresses that there are 
exactly $a$ different interpretations of the variable $y$ such that the formula $\varphi(y)$
is satisfied. This leads to the logic called $\FOunC$, first-order logic with unary counting
quantifiers. Similarly, by adding quantifiers that allow to count the number of 
$k$-tuples that satisfy a formula, one obtains the logic $\FOkaryC$, first-order logic with 
$k$-ary counting quantifiers. 
\\
In \cite{BIS} it was shown that, on finite ordered structures, $\FOunC$
with varying arithmetic predicates 
corresponds to the circuit complexity class $\TCzero$ with varying uniformity conditions.
\par
In a different line of research, \emph{pure} arithmetic is considered. There, the 
underlying structure is either the set of natural numbers with certain arithmetic predicates,
or initial segments of $\NN$ with arithmetic predicates --- and the signature contains 
nothing else but the arithmetic predicates. The aim is to investigate and
compare the expressive power of first-order logic with different arithmetic predicates.
Concerning $\NN$, detailed overviews can be found in \cite{Bes,Korec}; concerning initial
segments of $\NN$, we refer to \cite{EsbelinMore} and the references therein. One important
open question is whether the so-called class of \emph{rudimentary relations} is closed under 
counting, i.e., whether on initial segments of 
$\struc{\NN,+,\times}$ first-order logic is as expressive as $\FOunC$.
\par
The aim of the present paper is to
\begin{enumerate}[$\bullet$]
\item
  give an overview of what is known about the expressive power of first-order logic
  with different arithmetic predicates. The emphasis here lies on finite structures and
  initial segments of $\NN$ rather than $\NN$.
\item
  examine in detail the expressive power of first-order logic with counting quantifiers 
  and with 
  different arithmetic predicates, for finite structures as well as for {pure}
  arithmetic on $\NN$ and on initial segments of $\NN$.
  In particular, we will point out that on the (non-ordered) structure $\struc{\NN,\times}$ 
  the use of the logic $\FOunC$ does not make sense, since this logic lacks to have the 
  isomorphism property on $\struc{\NN,\times}$ and its initial segments. I.e., for
  $\struc{\NN,\times}$ and its initial segments the usual approach with two-sorted 
  structures would be more adequate. 
\item
  give a positive answer to the analogue of the above question on rudimentary relations, 
  for Presburger arithmetic $\struc{\NN,+}$ rather than $\struc{\NN,+,\times}$. I.e., 
  we will show that on $\struc{\NN,+}$ and its initial segments first-order logic is indeed
  as expressive as $\FOunC$. 
  As applications of this result we will obtain the failure of a particular version of the 
  so-called Crane Beach conjecture, we will obtain
  an easy proof of Ruhl's result \cite{Ruhl} that 
  reachability in finite graphs is not expressible in $\FOunC(+)$ and, similarly, 
  that connectivity of finite graphs is not definable in $\FOunC(+)$.   
\end{enumerate}%
Via communication with Leonid Libkin the author learned that the result on
Presburger arithmetic was independently discovered, but not yet published, by H.\ J.\ Keisler.
\\
Let us mention two
more papers that deal with unary counting quantifiers and with $\FO(+)$, 
respectively:
Benedikt and Keisler \cite{BenediktKeisler} \index{Benedikt, Michael}\index{Keisler, H.\ J.}
investigated several different kinds 
of unary counting quantifiers. Implicitly, they show that, under certain presumptions, 
such unary counting quantifiers can be eliminated (cf., Lemma~19 in the 
appendix of \cite{BenediktKeisler}). However, their result does not deal with 
Presburger arithmetic and its initial segments, and their proofs are
non-elementary, using non-standard models and hyperfinite structures.
Pugh \cite{Pugh} \index{Pugh, William}
deals with Presburger arithmetic $\struc{\ZZ,<,+}$ and counting 
quantifiers from a different point of view. He
presents a way of how a symbolic math package such as \emph{Maple} or \emph{Mathematica}
may compute symbolic sums of the form
\,$
\sum\setc{p(\vek{y},\vek{z})}{\vek{y}\in\ZZ\mbox{ and } \struc{\ZZ,<,+}\models\varphi(\vek{y},\vek{z})}
$,\,
where $p$ is a polynomial in the variables $\vek{y},\vek{z}$ and $\varphi$ is a
$\FO(<,+)$-formula. The 
$\FOkaryC$-formulas considered in the present paper
correspond to the simplest such sums in which the polynomial $p$ is the constant $1$.
\\ \parno
The present paper contains results of the author's dissertation \cite{Schweikardt_Diss}.
The paper is structured as follows:
Section~\ref{section:Preliminaries} fixes the basic notations concerning first-order
logic.
Section~\ref{sec:FO_Arithmetic} summarizes important properties of 
first-order logic with arithmetic predicates, concentrating on its ability and its
inability, respectively, to count cardinalities of certain sets.
Section~\ref{section:FO_Count} fixes the syntax and semantics of first-order 
logic with counting
quantifiers and exposes important properties of this logic.
In Section~\ref{section:PresburgerArithmetic} we show that Presburger arithmetic is closed
under unary counting quantifiers.
Section~\ref{section:Applications} points out some applications of the previous section's
result: We obtain the failure of a particular version of the Crane Beach conjecture, and 
we show that reachability and
connectivity of finite graphs are not expressible in first-order logic with unary counting
and addition.
Finally, Section~\ref{section:Conclusion} points out further questions and gives a 
diagram that visualizes the expressive
power of first-order logic with counting quantifiers and various arithmetic predicates.
\\ \parno
\textbf{Acknowledgements:} I want to thank Clemens Lautemann, Malika More, and 
Thomas Schwentick for helpful discussions on the subject of this paper. 
Especially the proof of 
Proposition~\ref{prop:Mult_IsomProperty} is partly due to them.




\section{Preliminaries}\label{section:Preliminaries}
%
%
\subsection{Basic Notations}
We use $\ZZ$ for the set of integers, $\NN\deff \set{0,1,2,\twodots}$
\index{N-@$\NN$}\index{Z@$\ZZ$}\index{Q1@$\QQ$}\index{R@$\RR$}%
for the set of natural numbers, and $\NNpos$ for the set of positive
natural numbers.
For $N\in\NN$ we write $\dom{N}$ \index{N@$\dom{N} = \set{0,\twodots,N}$}%
to denote the initial segment
$\set{0,\twodots,N}$ of $\NN$.
\par
For $a,b\in\ZZ$ we write \,$a\mid b$\, to express that $a$ divides
$b$, \index{$\equiv_n$t@$\mid$ \,($a\mid b$)}%
i.e., that $b = c\cdot a$ for some $c\in\ZZ$.
We write $\lcm\set{n_1,\twodots,n_k}$ to denote the 
least common multiple of $n_1,\twodots,n_k\in\NNpos$, i.e., to denote the 
smallest number in $\NNpos$ that is divided by $n_i$, for every $i\in\set{1,\twodots,k}$.
For $n\in\NNpos$ the symbol $\equiv_n$ \index{$\equiv_n$ ($a\equiv_n b$)}%
denotes the \emph{congruence relation modulo $n$}, i.e., for $a,b\in\ZZ$ we have $a\equiv_n b$ \,iff \,$n \mid a{-}b$. 
The relation $\equiv_n$ can be extended to rational numbers $r,s$
via \,$r\equiv_n s$\, iff \,$r{-}s= z\cdot n$\, for some $z\in\ZZ$. 
For a rational number $r$ we write $\abgerundet{r}$ to denote the
\index{$\abgerundet{r}$, $\aufgerundet{r}$}%
largest integer $\leq r$, and 
$\aufgerundet{r}$ for the smallest integer $\geq r$.
By $\lg(r)$ \index{lg@$\lg(N)$}%
we denote the logarithm of $r$ with respect to base $2$.
\par
By $\emptyset$ we denote the empty set, $|A|$ denotes the cardinality of a
set $A$, and $A^m \deff \setc{(a_1,\twodots,a_m)}{a_1,\twodots,a_m\in
  A}$ is the set of all $m$-tuples in $A$.
Depending on the particular context, we use $\vek{a}$ as abbreviation for a
sequence $a_1,\twodots,a_m$ or a tuple $(a_1,\twodots,a_m)$.
%
An $m$-ary relation $R$ on $A$ is a subset of $A^m$. 
Instead of $\vek{a}\in R$ we often write $R(\vek{a})$.
\subsection{Signatures, Structures, and Isomorphisms}
A \emph{signature} $\tau$ consists of (a possibly infinite number of) 
constant symbols, relation symbols, and function symbols. Each relation or function
symbol $S\in \tau$ has a fixed arity 
$\ar(S)\in\NNpos$.
Whenever we refer to some ``$R\in\tau$'' we implicitly assume that $R$ is a
\emph{relation} symbol. Analogously, ``$c\in\tau$'' 
means that $c$ is a {constant} symbol, and ``$f\in\tau$''
means that $f$ is a {function} symbol.
\par
A \emph{$\tau$-structure} \index{structure!$\tau$-structure}\index{tstructure@$\tau$-structure}
$\A = \struc{A,\tau^{\A}}$ consists of an arbitrary set $A$ which
is called the \emph{universe} of $\A$, and a
set $\tau^{\A}$ that contains
   an interpretation $c^{\A} \in A$ for each $c\in\tau$,
   an interpretation $R^{\A} \subseteq A^{\ar(R)}$ for each
   $R\in\tau$, and
   an interpretation $f^{\A} : A^{\ar(f)} \rightarrow A$ for each
    $f\in\tau$.
%
The structure $\A$ is called \emph{finite} iff its universe $A$ is
finite. 
\par
An \emph{isomorphism} $\pi$ between two $\tau$-structures 
$\A = \struc{A,\tau^{\A}}$ and
$\B = \struc{B,\tau^{\B}}$ is a bijective mapping $\pi:A\rightarrow B$ such that
   $\pi(c^{\A}) = c^{\B}$ (for each $c\in\tau$), \ 
   $R^{\A}(\vek{a})$  iff  $R^{\B}\big(\pi(\vek{a})\big)$ (for each
   $R\in\tau$ and all $\vek{a}\in A^{\ar(R)}$), \ and
   $\pi\big(f^{\A}(\vek{a})\big) = f^{\B}\big(\pi(\vek{a})\big)$ (for
   each $f\in\tau$ and all $\vek{a}\in A^{\ar(f)}$).
%
An \emph{automorphism} of $\A$ \index{automorphism}%
is an isomorphism between $\A$ and $\A$.

\subsection{First-Order Logic}
Let $\tau$ be a signature. 
We use $x_1,x_2,\ldots$ as variable symbols.
\emph{$\tau$-terms} are built from the variable symbols,
the constant symbols, and the function symbols in
$\tau$ in the following way: Each constant symbol in $\tau$ and each variable symbol is a
$\tau$-term, and if $t_1,\twodots,t_m$ are $\tau$-terms and $f$ is a
function symbol in $\tau$ of arity $m$, then $f(t_1,\twodots,t_m)$ is a $\tau$-term.
\emph{Atomic $\tau$-formulas} are $t_1{=}t_2$ and \index{formula}
$R(t_1,\twodots,t_{m})$, where $R\in\tau$ is of arity $m$ and
$t_1,\twodots,t_{m}$ are $\tau$-terms.
\par
\emph{First-order $\tau$-formulas}, for short: \emph{$\FO(\tau)$-formulas},\index{FO@$\FO$}
are built up as usual from the atomic $\tau$-formulas and
the logical connectives $\oder$, $\nicht$, the variable symbols
$x_1,x_2,\ldots$, and the existential quantifier $\exists$. 
As usual, we use \,$\forall x\,\varphi$ \,(respectively
\,$\varphi\und\psi$, \,$\varphi\impl\psi$, \,$\varphi\gdw\psi$) \,as
abbreviation for \,$\nicht\exists x\,\nicht\varphi$ \,(respectively
\,$\nicht(\nicht\varphi \oder \nicht\psi)$, \,$\nicht\varphi\oder\psi$,
\,$(\varphi\und\psi)\oder(\nicht\varphi\und\nicht\psi)$). 
\par
With $\free(\varphi)$ we denote the set of all \index{free@$\free(\varphi)$}
variables that occur free (i.e., not in the scope of some quantifier) in
$\varphi$.
Sometimes we write $\varphi(x_1,\twodots,x_m)$ to indicate that
$\free(\varphi)\subseteq\set{x_1,\twodots,x_m}$.
We say that $\varphi$ is a \emph{sentence} \index{sentence}
if it has no free variables. 
We say that $\varphi$ is \emph{quantifier free} \index{quantifier free} 
if there is no quantifier in $\varphi$ (i.e., $\varphi$ is a Boolean combination of
{atomic} $\tau$-formulas).
\\
If we insert additional relation, function, or constant symbols, e.g.,
$<$ and $+$, into a 
signature $\tau$, we simply write $\FO(\tau,<,+)$ instead of
$\FO(\tau\cup\set{<,+})$. 
%
\par
For a $\FO(\tau)$-sentence $\varphi$ and a $\tau$-structure $\A$ we
say that $\A$ models $\varphi$ and 
write $\A\models\varphi$ to indicate that $\varphi$ is satisfied when
interpreting each symbol in $\tau$ by its interpretation in $\tau^{\A}$.
For a $\FO(\tau)$-formula $\varphi(x_1,\twodots,x_m)$ and for interpretations
${a_1},\twodots,{a_m} \in A$ of the variables $x_1,\twodots,x_m$, we write
$\A\models\varphi({a_1},\twodots,{a_m})$ 
(or, equivalently, $\struc{\A,a_1,\twodots,a_m}\models\varphi({x_1},\twodots,{x_m})$)
to indicate that the
$(\tau\cup\set{x_1,\twodots,x_m})$-structure
$\struc{\A,{a_1},\twodots,{a_m}}$ models the
$\FO(\tau,{x_1,\twodots,x_m})$-sentence $\varphi$.
\par
It should be obvious that $\FO(\tau)$ has the \emph{isomorphism
  property},\index{isomorphism property} i.e.:
If $\pi$ is an isomorphism between two $\tau$-structures $\A$ and $\B$, if $\varphi(\vek{x})$ is a
$\FO(\tau)$-formula, and if $\vek{a}\in A$ is an interpretation of the variables $\vek{x}$, then
 \ $\A\models\varphi(\vek{a})$\, iff 
   \,$\pi(\A)\models\varphi\big(\pi(\vek{a})\big)$.
\par
A relation ${R}\subseteq A^m$ is called \emph{$\FO(\tau)$-definable} in $\A$ 
if there is a $\FO(\tau)$-formula $\varphi(x_1,\twodots,x_m)$ such that
\,${R} = 
 \setc{({a_1},\twodots,{a_m})\in A^m}{\A\models\varphi({a_1},\twodots,{a_m})}$.
Ac\-cordingly, a function ${f}:A^m\rightarrow A$ and an element $a\in A$ are called
$\FO(\tau)$-definable in $\A$ if the corresponding relations
\,${R_f}\deff
 \setc{\big({a_1},\twodots,{a_m},
   {f}({a_1},\twodots,{a_m})\big)}{({a_1},\twodots,{a_m})\in A^m}$\,
and
\,${R_a}\deff\set{{a}}$\,
are \emph{$\FO(\tau)$-definable} in $\A$. 
\par
We say that two $\FO(\tau)$-formulas $\varphi(\vek{x})$ and $\psi(\vek{x})$ 
are \emph{equivalent over $\A$} if, for all interpretations ${\vek{a}}\in A$ of the 
variables $\vek{x}$, we have
\;\(
   \A \models \varphi({\vek{a}})
   \, \mbox{ iff } \,
   \A \models \psi({\vek{a}})
\).
Accordingly, if $\K$ is a class of $\tau$-structures, we say that
$\varphi(\vek{x})$ and $\psi(\vek{x})$ are \emph{equivalent over $\K$}, if they are
equivalent over every structure $\A\in\K$.
%
%




\section{First-Order Logic with Arithmetic}\label{sec:FO_Arithmetic}
In this section we summarize important properties of first-order logic with 
arithmetic and we point out the correspondence between first-order logic with arithmetic and 
\emph{circuit complexity} on the one hand and \emph{rudimentary relations} on the other hand.
\subsection{Arithmetic}
In this paper we consider the following \emph{arithmetic predicates} on $\NN$ and on initial 
segments $\dom{N}$ of $\NN$: 
\begin{enumerate}[$\bullet$]
\item
  the binary \emph{linear ordering} predicate $<$,
\item
  the ternary \emph{addition} predicate $+$,
  consisting of all triples $(x,y,z)$ such that $x+y=z$,
\item
  the ternary \emph{multiplication} predicate $\times$,
  consisting of all triples $(x,y,z)$ such that \,$x\cdot y=z$,
\item
  the ternary \emph{exponentiation} predicate $\Exp$, \index{Exp@$\Exp$} 
  consisting of all triples $(x,y,z)$ such that $x^y=z$,
\item
  the binary \emph{Bit} predicate $\Bit$, \index{Bit@$\Bit$ predicate} 
  consisting of all tuples $(x,y)$ such that the $y$-th
  bit in the binary representation of $x$ is $1$, i.e., 
  $\abgerundet{\frac{x}{2^y}}$ is odd,
\item
  the unary \emph{square numbers} predicate $\Squares$, \index{Squares@$\Squares$}
  consisting of all numbers $n^2$, for all $n\in\NN$.
\end{enumerate}%
When speaking of \emph{arithmetic on finite structures} we consider a set
$\frak{A}$ of arithmetic predicates. 
Furthermore, we consider
arbitrary signatures $\tau$ and all $\tau$-structu\-res whose
universe is an initial segment of $\NN$. Given such a $\tau$-structure 
$\A = \struc{\dom{N},\tau^{\A}}$
we enrich $\A$ by the arithmetic predicates in $\frak{A}$. 
I.e., we move over 
to the \,$(\tau\cup\frak{A})$-structure 
\,$\struc{\A,\frak{A}} \deff \struc{\dom{N},\tau^{\A},\frak{A}^N}$, \,
where $\frak{A}^N$ is the collection of the relations $P^N\deff P\cap\dom{N}^{\ar(P)}$, 
for all $P\in\frak{A}$.
Usually we will suppress the superscript $N$ and simply write 
 $\frak{A}$ instead of $\frak{A}^N$ and $P$ instead of $P^N$. 
\\
In contrast to arithmetic on finite structures, \emph{pure arithmetic} means that we 
restrict our attention to structures where the signature $\tau$ is \emph{empty}.
I.e., we only consider the structure $\struc{\NN,\frak{A}}$ and the structures
$\struc{\dom{N},\frak{A}^N}$, for all $N\in\NN$.
\par
To compare the expressive power of different sets of arithmetic predicates, we fix the
following notation.
\begin{definition_ohne}
Let $\frak{A}_1$ and $\frak{A}_2$ be classes of arithmetic predicates, i.e., subsets of
$\set{<,+,\times,\Exp,\Bit,\Squares}$.
\begin{enumerate}[(a)\ ]
\item
The statement
{\bf ``$\bs{\FO(\frak{A}_1)\subseteq\FO(\frak{A}_2)}$ on $\bs{\NN}$''}
has the following precise meaning:
For every $\FO(\frak{A}_1)$-formula $\varphi_1(\vek{x})$ there is a $\FO(\frak{A}_2)$-formula
$\varphi_2(\vek{x})$ such that
\ ``$
  \struc{\NN,\frak{A}_1}\models \varphi_1(\vek{a})
  \ \mbox{ iff } \
  \struc{\NN,\frak{A}_2}\models \varphi_2(\vek{a})
$'' \ 
is true for all interpretations $\vek{a}\in\UU$ of the 
variables $\vek{x}$.
\item
The statement
{\bf ``$\bs{\FO(\frak{A}_1)\subseteq\FO(\frak{A}_2)}$ on initial segments of $\bs{\NN}$''}
has the following precise meaning:
For every $\FO(\frak{A}_1)$-formula $\varphi_1(\vek{x})$ there is a $\FO(\frak{A}_2)$-formula
$\varphi_2(\vek{x})$ such that 
\ ``$ 
  \struc{\dom{N},\frak{A}_1}\models \varphi_1(\vek{a})\allowbreak
  \ \mbox{ iff } \
  \struc{\dom{N},\frak{A}_2}\models \varphi_2(\vek{a})
$'' \ 
is true for all $N\in\NNpos$ and all interpretations $\vek{a}\in\dom{N}$ of the 
variables $\vek{x}$.
\item
The statement
{\bf ``$\bs{\FO(\frak{A}_1)\subseteq\FO(\frak{A}_2)}$ on finite structures''}
has the following precise meaning:
For every signature $\tau$ and
every $\FO(\frak{A}_1,\tau)$-formula $\varphi_1(\vek{x})$ there is a $\FO(\frak{A}_2,\tau)$-formula
$\varphi_2(\vek{x})$ such that
\ ``$
  \struc{\dom{N},\frak{A}_1,\tau^{\A}}\models \varphi_1(\vek{a})
  \ \mbox{ iff } \ 
  \struc{\dom{N},\frak{A}_2,\tau^{\A}}\models \varphi_2(\vek{a})
$'' \ 
is true for all $N\in\NNpos$, 
for all $\tau$-structures \,$\A =\struc{\dom{N},\tau^{\A}}$\,  
and for all interpretations $\vek{a}\in \dom{N}$ of the variables $\vek{x}$.
\mbox{}\fertig
\end{enumerate}
\end{definition_ohne}%
\subsection{Expressive Power}\label{subsection:expressive_power} 
The expressive power of first-order logic with 
arithmetic predicates $<$, $+$, $\times$, etc.\ is by now well understood:
{
 \bf
\begin{displaymath}
 \begin{array}{l}
 \bs{\FO(<)\ \subsetneq\  \FO(+)\  \subsetneq\  \FO(+,\times)\qquad \mbox{and}}
 \\[1ex]
 \bs{
    \FO(+,\times) 
  \ =\  \FO(<,\times)  
  \ =\  \FO(\Bit) 
  \ =\  \FO(+,\Squares)
  }
 \\
 \bs{
    \hphantom{\FO(+,\times)}
  \ =\ \FO(<,+,\times,\Exp,\Bit,\Squares)}
 \\[1ex] 
 \mbox{\normalsize \bf on initial segments of $\NN$ 
 (and on finite structures and on $\NN$).}
 \end{array}
\end{displaymath}
}%
%
More precisely:
\begin{enumerate}[$\bullet$]
\item
 {\bf$\bs{\FO(<) \subsetneq \FO(+)}$} \ is true, because, on the one hand, ``$<$'' can be
 expressed using ``$+$'', and on
 the other hand, there is a $\FO(+)$-formula, but no $\FO(<)$-formula which expresses that the
 cardinality of the underlying universe is even 
 (cf., e.g., the textbook \cite[Example~2.3.6]{EbbinghausFlum}).
\item
 {\bf$\bs{\FO(+) \subsetneq \FO(+,\times)}$} \ is true, because, there is
 a $\FO(+,\times)$-formula, but no $\FO(+)$-formula which expresses that the
 cardinality of the underlying universe is a prime number. This is a direct consequence of 
 the Theorem of Ginsburg and Spanier \index{Ginsburg, Seymour}\index{Spanier, Edwin H.}
 which states that the spectra of
 $\FO(+)$-sentences are \emph{semi-linear}, i.e., for every $\FO(+)$-sentence $\varphi$ there
 are numbers $p,N_0\in\NN$ such that for every $N> N_0$ we have $\struc{\dom{N},+}\models
 \varphi$ \ iff \ $\struc{\dom{N{+}p},+}\models\varphi$. 
 A proof of the Theorem of Ginsburg and Spanier, based on Presburger's quantifier elimination, can be found in the textbook 
 \cite[Theorem~4.10]{Smorynski}; an Ehrenfeucht-Fra\"\i{}ss\'{e} game proof 
 is given in \cite[Corollary~8.5]{Schweikardt_Diss}.
\item
 {\bf $\bs{\FO(+,\times) = \cdots = \FO(<,+,\times,\Exp,\Bit,\Squares)}$} \ is true
 because of the following:
\end{enumerate}%
\begin{theorem_ohne}\label{theorem:arithmetic}
There is
\begin{enumerate}[(a)]
 \item 
  a $\FO(\Bit)$-formula $\varphi_{<}(x,y)$, such that for every $N\in\NNpos$ and all
  assignments $\bs{x},\bs{y}\in\dom{N}$ of the variables
  ${x},{y}$, we have \ 
  $\struc{\dom{N},\Bit} \models \varphi_{<}(\bs{{x}},\bs{{y}})$ \ iff\ 
  \ $\bs{x} < \bs{y}$.
 \item 
  a $\FO(\Bit)$-formula $\varphi_{+}(x,y,z)$, such that for every $N\in\NNpos$ and all
  assignments $\bs{x},\bs{y},\bs{z}\in\dom{N}$ of 
  ${x},{y},z$, we have \ 
  $\struc{\dom{N},\Bit} \models \varphi_{+}(\bs{{x}},\bs{{y}},\bs{z})$ \ iff\ 
  \ $\bs{x} + \bs{y} = \bs{z}$.
 \item 
  a $\FO(\Bit)$-formula $\varphi_{\times}(x,y,z)$, such that for every $N\in\NNpos$ and all
  assignments $\bs{x},\bs{y},\bs{z}\in\dom{N}$ of 
  ${x},{y},z$, we have \ 
  $\struc{\dom{N},\Bit} \models \varphi_{\times}(\bs{{x}},\bs{{y}},\bs{z})$ \ iff\ 
  \ $\bs{x} \times \bs{y} = \bs{z}$.
 \item 
  a $\FO(<,\times)$-formula $\varphi_{\Bit}(x,y)$, such that for every $N\in\NNpos$ and all
  assignments $\bs{x},\bs{y}\in\dom{N}$ of the variables
  ${x},{y}$, we have \ 
  $\struc{\dom{N},<,\times} \models \varphi_{\Bit}(\bs{{x}},\bs{{y}})$ \ iff\ 
  \ $\Bit(\bs{x},\bs{y})$.
 \item 
  a $\FO(+,\times)$-formula $\varphi_{\Exp}(x,y,z)$, such that for every $N\in\NNpos$ and all
  assignments $\bs{x},\bs{y},\bs{z}\in\dom{N}$ of 
  ${x},{y},{z}$, we have \
  $\struc{\dom{N},+,\times} \models \varphi_{\Exp}(\bs{{x}},\bs{{y}},\bs{{z}})$ \ iff\ 
  \ $\bs{x} = \bs{y}^{\bs{z}}$.
 \item 
  a $\FO(+,\Squares)$-formula $\varphi_{\times}(x,y,z)$, such that for every $N\in\NNpos$ and 
  all assignments $\bs{x},\bs{y},\bs{z}\in\dom{N}$ of 
  ${x},{y},{z}$, we have \ 
  $\struc{\dom{N},+,\Squares} \models \varphi_{\times}(\bs{{x}},\bs{{y}},\bs{{z}})$ \ iff\ 
  \ $\bs{x}\times \bs{y} = \bs{z}$.
  \mbox{}\fertig
\end{enumerate}
\end{theorem_ohne}%
\begin{proof_ohne}
The proofs of the parts \emph{(a)--(e)} are very involved.\\
Part \emph{(a)} was shown by Dawar et al.\ in \cite{DDLW}.\index{Dawar, Anuj} \\
$\FO(<,\Bit)$-formulas for \emph{(b)} and
\emph{(c)} are outlined in the textbook \cite{Immerman}, where
also the construction of a
$\FO(<,+,\times)$-formula for \emph{(d)} is described. 
Troy Lee \index{Lee, Troy}
observed in an email note to Immerman \index{Immerman, Neil} and 
Barrington \index{Barrington, David A.\ Mix} that in this construction 
all the uses of $+$ can be replaced using $<$, and thus $<$ and
$\times$ suffice to express $\Bit$.\\
Part \emph{(e)} was shown by Bennet \index{Bennet, J.\ H.}
in \cite{Bennet} (see also
Lindell's email note \cite{LindellEmail}).\index{Lindell, Steven}
\\
The proof of part \emph{(f)} is not so difficult: \par
\emph{Step 1} is to construct a $\FO(+,\Squares)$-formula $\psi(u,v)$ expressing that
$u^2=v$.
Here, one can make use of the equation \,$(u-1)^2 = u^2 - 2u + 1$\, which gives us that
$u^2=v$ is valid \ if and only if 
\[
\begin{array}{cl}
 - & \mbox{$v$ is a square number, i.e., $\Squares(v)$, \ and}
 \\[1ex]
 - & \mbox{($u=0$\, and \,$v=0$) \ or \ ($u=1$\, and \,$v=1$) \ or }
 \\[1ex]
 - & \mbox{for the number $w$ that is the predecessor of $v$ in the set $\Squares$ we} 
 \\
   & \mbox{have that \,$w = v - 2u + 1$.}
\end{array}
\]
It is straightforward to express this by a $\FO(+,\Squares)$-formula $\psi(x,y)$.
\par
\emph{Step 2} is to construct a $\FO(+,\Squares)$-formula $\varphi'_{\times}(x,y,z)$ 
expressing that $x\times y = z$ for numbers $x,y$ of size at most $\sqrt{N}$ (when considering
the universe $\set{0,\twodots,N}$).
Here, one can make use  of the equation \,$(x-y)^2 = x^2 - 2xy + y^2$\, which gives us that
$x\times y=z$ \ if and only if \ the equation $w = u - 2z + v$ is true for the numbers
$u\deff x^2$, $v\deff y^2$, and $w\deff (x{-}y)^2$.
Using the formula $\psi$ from Step~1, it is straightforward to express this by 
a $\FO(+,\Squares)$-formula $\varphi'_{\times}(x,y,z)$. 
Note that this formula defines the multiplication $\times$ only for numbers $x,y$ of size
at most $\sqrt{N}$, where $N$ is the maximum element in the universe.
\par
\emph{Step 3} is to lift the multiplication from numbers of size up to $\sqrt{N}$ to numbers
of size up to $N$. Such a lifting is proved in \cite[Lemma\;1\,(ii)]{Lynch_NTIME}. 
The details are similar to the details in Step 2 of the 
proof of Theorem~\ref{theorem:d-arithmetic}\,(d) in the appendix of the present paper. 
The basic idea is the following:
\begin{enumerate}[1.]
\item 
  For numbers $x\in\set{0,\twodots,N}$ use the $(M{+}1)$-ary decomposition
  $x = x_1\cdot (M{+}1) + x_0 $, \,where $x_1,x_0 \leq M$\, and
  \,$M \deff \lfloor\sqrt{N}\rfloor$.  
\item 
  Show that this decomposition is definable via a
  $\FO(+,\Squares)$-formula 
  $\chi(x,x_1,x_0)$.
\item
  Use $\varphi'_{\times}$ to construct a formula
  \,{$\varphi''_{\times}(x_1,x_0,y_1,y_0,z_1,z_0)$}\, that defines the 
  multiplication for the $(M{+}1)$-ary decompositions of numbers $x,y,z$.  
\end{enumerate}
Finally this leads to the desired $\FO(+,\Squares)$-formula that defines 
multiplication of numbers of size up to $N$. Hence, the proof sketch for 
part \emph{(f)} of Theorem~\ref{theorem:arithmetic} is complete.
\mbox{ } \qed
\end{proof_ohne}%
\\
\parno
It is easy to see that ``$<$'' cannot be expressed using ``$\times$'' alone, i.e.,
\begin{quote}\bf
 $\bs{\FO(\times)\, \subsetneq\, \FO(<,\times)}$ \ 
 on initial segments of $\bs{\NN}$ 
 (and also on finite structures and on $\bs{\NN}$).
\end{quote}
To see this, let $\A$ be either the structure $\struc{\NN,\times}$ or some initial segment
$\struc{\dom{N},\times}$. For the sake of contradiction, assume that there is a
$\FO(\times)$-formula $\varphi_<(x,y)$ expressing that $a{<}b$, for all interpretations 
$a,b\in A$ of the variables $x,y$.
The \emph{isomorphism property} of $\FO(\times)$ thus implies, for every automorphism $\pi$ of $\A$, that 
$\pi(a){<}\pi(b)$ iff $a{<}b$. Hence, the identity function on $A$ is the only
automorphism of $\A$. 
\\
The contradiction now follows from the fact that $\struc{\NN,\times}$ and
also most initial segments $\struc{\dom{N},\times}$ do have automorphisms different from the
identity function:
Indeed, over $\NN$, the role of any two
different prime numbers $p$ and $q$ is interchangeable. I.e., the following
mapping $\pi_{p\gdw q}$ is an automorphism of $\struc{\NN,\times}$:
$\pi_{p\gdw q}$ is determined via
\,$\pi_{p\gdw q}\,(a\times b) \,=\, \pi_{p\gdw q}\,(a)\times
\pi_{p\gdw q}\,(b)$\, for all $a,b\in\NNpos$,\, and,
for all prime numbers $r$,
{ \small
\[
   \pi_{p\gdw q}\,(r) \quad \deff\quad
   \left\{
     \begin{array}{ll}
        q & \mbox{ if \ } r=p \\
        p & \mbox{ if \ } r=q \\
        r & \mbox{ if \ $r\neq p,q$ is a prime.}
     \end{array}
   \right.
\]
}%
Moreover, if $p$ and $q$ are prime numbers $>\frac{N}{2}$, then $\pi_{p\gdw q}$ can even be viewed
as an automorphism of the initial segment $\struc{\dom{N},\times}$. In fact, $\pi_{p\gdw q}$ 
leaves all elements in $\dom{N}$ fixed except for $p$ and $q$. For example, $\pi_{2\gdw 3}$ is
an automorphism of $\struc{\dom{3},\times}$, and $\pi_{5\gdw 7}$ is
an automorphism of $\struc{\dom{8},\times}$. Moreover, from results in
number theory (cf., e.g., \cite[Problem~17 in Chapter~13]{Zahlentheorie}) we
know that for any large enough $N$ there are prime numbers $p,q$ with
$\frac{N}{2}< p<q\leq N$. 
\par
What we have seen is that there is no $\FO(\times)$-formula $\varphi_<(x,y)$ such that
\ ``$\struc{\dom{N},\times}\models \varphi_<(a,b)$ \,iff\, $a{<}b$'' \ is true for all
$N\in\NNpos$ and all $a,b\in\dom{N}$.
It is remarkable, however, that ``$<$'' is indeed $\FO(\times)$-definable on numbers of size 
up to $\sqrt{N}$: 
\begin{lemma_mit}[Folklore]\label{lemma:kleiner_auf_Anfangsstueck} 
There is a $\FO(\times)$-formula $\varphi'_<(x,y)$ which defines 
``$<$'' on numbers of size up to $\sqrt{N}$.
I.e., for all $N\in\NNpos$ and all interpretations $a,b\in\dom{N}$ of the
variables $x,y$, we have \ $\struc{\dom{N},\times}\models \varphi'_<(a,b)$ \ iff \  
$a< b \leq \sqrt{N}$.
\end{lemma_mit}%
\begin{proof_mit}
The $\FO(\times)$-formula $\varphi'_<(x,y)$ is defined via 
\[
  \exists z\; x{\times}x{=}z\ \und\ \exists z\; y{\times}y{=}z\ \und\ 
  \exists u\ \big(\, \big(\exists v\; x{\times} u{=}v\big) \,\und\, 
              \big(\nicht\exists v' \; y{\times} u{=}v' \big)\,\big). 
\]
For the ``only if'' direction let $a,b\in\dom{N}$ such that 
$\struc{\dom{N},\times}\models \varphi'_<(a,b)$.
Clearly, the first two conjunctions of $\varphi'_<$ ensure that  
$a,b\leq \sqrt{N}$. The third conjunction ensures that there is some $u\in\dom{N}$ such that
$a{\times}u \leq N$ and $b{\times}u > N$, and hence, in particular  $a<b$.
\\
For the ``if'' direction let $a < b\leq \sqrt N$. In particular,
$a{\times}a\leq N$ and 
$b{\times}b\leq N$, and hence the first two conjunctions of $\varphi'_<$ are satisfied.
Choose $u\in\dom{N}$ maximal such that $a{\times}u\leq N$. In particular, $u\geq a$, 
and there is some 
$r$ with $0\leq r<a\leq u$ such that $N= a{\times}u + r$. Since $b \geq a{+}1$ we thus have
$b{\times}u \geq (a{+}1)\times u = a{\times}u + u > a{\times}u + r = N$. Hence, also the third
conjunction in $\varphi'_<$ is satisfied.
\end{proof_mit}%
\\
\parno
When considering an initial segment $\dom{N}$, the relations $<$, $+$,
$\times$, $\Bit$ can a priori speak only about numbers of size at most
$N$.
This can be improved up to $N^d$ (for any fixed $d\in\NNpos$) by using
$(N{+}1)$-ary representations of numbers:
We use a $d$-tuple $\vek{x}\deff (x_{d-1},\twodots,x_0)\in(\dom{N})^d$
to represent the number $\sum_{i=0}^{d-1}x_i (N{+}1)^i$. The following
Theorem~\ref{theorem:d-arithmetic} shows that  
\begin{quote}\bf  
the $\bs{d}$-tuple versions of $\bs{<}$, $\bs{+}$, $\bs{\times}$, and 
$\bs{\Bit}$, respectively, are first-order definable on initial segments of
$\bs{\NN}$.  
\end{quote}      
This fact has been observed and used in various places, e.g., 
\cite{Harrow,Atserias}.  
\begin{theorem_ohne}[Folklore]\label{theorem:d-arithmetic}
For every $d\in\NNpos$ there is
\begin{enumerate}[(a)]
\item 
  a $\FO(<)$-formula $\varphi_{<}^{d}(x_{d-1},\twodots,x_0,\allowbreak
  y_{d-1},\twodots,y_0)$, such that for every $N\in\NNpos$ and all
  assignments $\bs{\vek{x},\vek{y}}\in\dom{N}$ of the variables
  $\vek{x},\vek{y}$, we have 
  $\struc{\dom{N},<} \models \varphi_{<}^{d}(\bs{\vek{x}},\bs{\vek{y}})$ \ iff\\ 
  \(
    \sum_{i=0}^{d-1} \bs{x_i}\,(N{+}1)^i \ < \ \ \sum_{i=0}^{d-1}
    \bs{y_i}\,(N{+}1)^i
  \).
\item 
  a $\FO(+)$-formula $\varphi_{+}^{d}(x_{d-1},\twodots,x_0,\allowbreak
  y_{d-1},\twodots,y_0,\allowbreak z_{d},\twodots,z_0)$, such that for every $N\in\NNpos$ and all
  assignments $\bs{\vek{x},\vek{y},\vek{z}}\in\dom{N}$ of the variables
  $\vek{x},\vek{y},\vek{z}$, we have \\ 
  $\struc{\dom{N},<,+} \models
  \varphi_{+}^{d}(\bs{\vek{x}},\bs{\vek{y}},\bs{\vek{z}})$ \ iff\\ 
  \(
    \sum_{i=0}^{d-1} \bs{x_i}\,(N{+}1)^i \ +\ \sum_{i=0}^{d-1} \bs{y_i}\,(N{+}1)^i \ = \ 
        \sum_{i=0}^{d} \bs{z_i}\,(N{+}1)^i   
  \).   
\item 
  for every fixed $n\in\NNpos$, a $\FO(+)$-formula $\varphi_{\equiv_n}^{d}(x_{d-1},\twodots,x_0,\allowbreak
  y_{d-1},\twodots,y_0)$, such that for every $N\in\NNpos$ and all
  assignments $\bs{\vek{x},\vek{y}}\in\dom{N}$ of the variables
  $\vek{x},\vek{y}$, we have \ 
  $\struc{\dom{N},<,+} \models
  \varphi_{\equiv_n}^{d}(\bs{\vek{x}},\bs{\vek{y}})$ \ iff \
  \(
    \sum_{i=0}^{d-1} \bs{x_i}\,(N{+}1)^i \ \equiv_n \ \ \sum_{i=0}^{d-1} \bs{y_i}\,(N{+}1)^i 
  \).  
\item 
  a $\FO(+,\times)$-formula $\varphi_{\times}^{d}(x_{d-1},\twodots,x_0,\allowbreak y_{d-1},
  \twodots,y_0,\allowbreak z_{2d-1},\twodots,z_0)$, such that for every $N\in\NNpos$ and all
  assignments $\bs{\vek{x},\vek{y},\vek{z}}\in\dom{N}$ of the variables
  $\vek{x},\vek{y},\vek{z}$, we have \ 
  $\struc{\dom{N},+,\times} \models
  \varphi_{\times}^{d}(\bs{\vek{x}},\bs{\vek{y}},\bs{\vek{z}})$ \ iff\\ 
  \(
    \sum_{i=0}^{d-1} \bs{x_i}\,(N{+}1)^i \ \times \ \sum_{i=0}^{d-1} \bs{y_i}\,(N{+}1)^i \ =\
        \sum_{i=0}^{2d-1} \bs{z_i}\,(N{+}1)^i 
  \).  
\item 
   a $\FO(\Bit)$-formula $\varphi^{\Bit}_{d}(x_{d-1},\twodots,x_0,y)$,
   such that for every $N\in\NNpos$ and all
   assignments $\bs{\vek{x},{y}}\in\dom{N}$ of the variables
   $\vek{x},{y}$, we have 
   $\struc{\dom{N},\Bit} \models \varphi_{\Bit}^{d}(\bs{\vek{x}},\bs{y})$ \ iff\\ 
  \(
     \mbox{the $\bs{y}$-th bit in the binary representation of \ } \sum_{i=0}^{d-1}
     \bs{x_i}\,(N{+}1)^i  \mbox{\;\ is 1.}
  \). 
  \mbox{ }\fertig
\end{enumerate}
\end{theorem_ohne}%
The proof of Theorem~\ref{theorem:d-arithmetic} is straightforward
but tedious. For the sake of completeness
--- since the author does not know references that contain
complete proofs of all parts of this theorem --- a proof is given in the appendix.
%
%
%
\subsection{Counting vs.\ Arithmetic on Finite Structures}\label{section:Preliminaries:Arithmetic:Finite}
There is a close connection between arithmetic on finite structures and circuit complexity.
\index{circuit complexity}
A concise overview of circuit complexity can be found in \cite{Allender}.
The complexity class $\ACzero$ \index{AC0@$\ACzero$} 
consists of all problems solvable by
polynomial size, constant depth circuits of AND, OR, and NOT gates of unbounded
fan-in. It was shown in \cite{BIS} that, for ordered structures
over arbitrary signatures $\tau$, logtime-uniform $\ACzero$ is exactly the class of
all problems definable in $\FO(+,\times,\tau)$. 
It is a deep result of \cite{Ajtai,FSS} that \index{Parity@$\Parity$} 
\begin{eqnarray*}
  \Parity & \deff &
  \setc{\,\struc{\dom{N},<,\bs{Y}}\,}{\,N\in\NNpos,\ \bs{Y}\subseteq\dom{N},\,\ |\bs{Y}| \mbox{
  is even}\,}
\end{eqnarray*}
does not belong to $\ACzero$, and hence is not definable in
$\FO(+,\times,Y)$.
This is known even for non-uniform $\ACzero$, which translates to
$\FO(\Arb,Y)$, where $\Arb$ is the collection of \emph{arbitrary}, i.e.\ all,
built-in predicates on initial segments of $\NN$.
From \cite{FKPS,DeGuSh} we also know that, for any $\varepsilon > 0$,
$\FO(\Arb,Y)$ cannot count cardinalities of sets up to size
$N^{\varepsilon}$:
\begin{theorem_mit}[$\bs{\FO(\Arb)}$ cannot count on finite structures]\label{theorem:non-counting-ability}\mbox{ }\\
Let $\varepsilon >0$. There is no $\FO(\Arb,Y)$-formula
$\chi_{\Count}(x,Y)$ such that
\begin{eqnarray*}
   \struc{\,\dom{N},\,\Arb,\,\bs{x},\,\bs{Y}\,}\ \models\ \chi_{\Count}(x,Y)
 & \mbox{ iff }
 & \bs{x} = |\bs{Y}| \leq N^{\varepsilon}
\end{eqnarray*}
is true for all \,$N\in\NNpos$,\, all \,$\bs{Y}\subseteq \dom{N}$,\, and all
\,$\bs{x}\in\dom{N}$.
\end{theorem_mit}%
\\
However, it was shown in \cite{FKPS,DeGuSh,AjtaiBenOr} that, for
any $c\in\NNpos$, $\FO(+,\times,Y)$ can indeed count cardinalities of sets
up to size $(\lg N)^c$: \index{counting capability}
\begin{theorem_mit}[Polylog Counting Capability of
  $\bs{\FO(+,\times)}$]\label{theorem:polylog-counting-ability}\mbox{ }\\
For every $c\in\NNpos$ there is a $\FO(+,\times,Y)$-formula
$\chi_{\Count}^c(x,Y)$ such that
\begin{eqnarray*}
   \struc{\,\dom{N},\,+,\,\times,\,\bs{x},\,\bs{Y}\,}\ \models\ \chi_{\Count}^c(x,Y)
 & \mbox{ iff }
 & \bs{x} = |\bs{Y}| \leq (\lg N)^{c}
\end{eqnarray*}
is true for all \,$N\in\NNpos$,\, all \,$\bs{Y}\subseteq \dom{N}$,\, and all
\,$\bs{x}\in\dom{N}$.
\end{theorem_mit}%
A self-contained, purely logical proof of this theorem can be found in
\cite{DLM}.
\subsection[Counting vs.\ Pure Arithmetic on Initial Segments of $\NN$]{Counting vs.\ Pure Arithmetic on Initial Segments of $\bs{\NN}$}\label{section:Preliminaries:Arithmetic:Pure}
\raus{
\begin{definition_mit}[Spectrum $\Spectrum$]\index{spectrum}\index{Spec@$\Spectrum$}
Let $\Num$ be a collection of built-in predicates on initial segments of $\NN$. 
The spectrum $\Spectrum(\varphi)$ of a $\FO(\Num)$-sentence
$\varphi$ is the set of all $N\in\NNpos$ for which \,$\struc{\dom{N},\Num}\models\varphi$.
\end{definition_mit}%
}
There is a direct correspondence between \emph{pure arithmetic}
$\FO(+,\times)$ on initial segments of $\NN$ and \emph{bounded
  arithmetic} $\Delta_0$ on $\NN$. \index{bounded arithmetic}
$\Delta_0$ is the class of all $\FO(+,\times)$-formulas in which
quantified variables are bounded by other variables via \,$\exists
x\,(x{\leq}y \und \ldots)$.
The $\Delta_0$-definable relations in $\NN$ are called the
\emph{rudimentary relations}. \index{rudimentary relation}
A recent overview of this line of research can be found in
\cite{EsbelinMore}, where it is also pointed out that
there is a precise correspondence between
\begin{enumerate}[(1.)\ ]
 \item the $\FO(+,\times)$-definable spectra (the \emph{spectrum} of a 
 $\FO(+,\times)$-sentence $\varphi$ is the set of all $N\in\NNpos$ such that 
 $\struc{\dom{N},+,\times}\models\varphi$),
 \item the unary rudimentary relations,
 \item the linear hierarchy $\LinH$, \ and \index{linear hierarchy}\index{LINH@$\LinH$}
 \item the string languages definable in monadic second
       order logic $\MSO(+)$. \index{MSO@$\MSO$} 
\end{enumerate}
\raus{
The correspondence between (1.) and (2.) is straightforward:
The spectrum of a $\FO(+,\times)$-sentence $\varphi$ is exactly the
unary rudimentary relation on $\NN$ defined by the $\Delta_0$-formula
$\varphi'(y)$ obtained from $\varphi$ by relativizing all
quantifications in $\varphi$ to numbers $\leq y$ (where $y$ is a new
variable not used in $\varphi$).
\\
The correspondence between (2.), (3.), and (4.) was established in
\cite{Wrathall,MoreOliveMSO}.
There, it was shown that a set $R\subseteq \NN$ is rudimentary if and
only if the set of $\set{\letter{1},\letter{2}}$-strings which are dyadic\footnote{One
  uses the \emph{dyadic} rather than the \emph{binary} representation
  because the dyadic representation constitutes a \emph{bijection} between
  strings and numbers, whereas in the binary representation the
string $s$ represents the same number as the strings $\letter{0}s$, $\letter{00}s$, etc.}
representations of elements in $R$, i.e., the set 
\begin{eqnarray*}
   \textit{dyadic}(R) & \deff &
   \setc{\,w\in\set{\letter{1},\letter{2}}^*}{w= w_{n}\cdots w_0 \,\mbox{ and }\,
   \sum_{i=0}^{n}w_i{\cdot} 2^i\;\in \,R\,}
\end{eqnarray*}
belongs to the linear hierarchy $\LinH$ which, in turn, is exactly the
set of strings definable in monadic second-order logic $\MSO(+)$.
}
Researchers concerned with rudimentary relations have developed clever
encoding techniques that expose the expressive power of bounded
arithmetic. For example, the exponentiation relation
\,$x{=}y^z$\, was proved to be rudimentary (and hence
$\FO(+,\times)$-definable on initial segments of $\NN$) already in
1962 by Bennet \cite{Bennet}. \index{Bennet, J.\ H.}
Furthermore, Theorem~\ref{theorem:d-arithmetic}
corresponds to Harrow's result \cite{Harrow} \index{Harrow, Keith}
that, for expressing
rudimentary relations, one may make use of polynomially bounded
quantification such as \,$\exists x\,(x{\leq}y^d \und\ldots)$.
Esbelin and More \cite{EsbelinMore} \index{Esbelin, Henri-Alex}\index{More, Malika}
developed a toolbox that allows to express
certain primitive recursive functions by $\Delta_0$-formulas.
\par
On the other hand, hardly any tools are known which enable us to prove
that some relation is \emph{not} rudimentary. 
According to \cite{EsbelinMore,ParisWilkie} it is still open whether
the rudimentary relations are closed under counting. Translated into
the setting used in the present paper, this corresponds to the
following:
\begin{question_mit}\label{question:counting_plus_mal}
Is there, for every $\FO(+,\times)$-formula $\varphi(y,\vek{z})$, a 
$\FO(+,\times)$-formula $\chi(x,\vek{z})$ such that
\begin{eqnarray*}
   \struc{\dom{N},+,\times}\,\models\,\chi(\bs{x},\bs{\vek{z}})
 & \mbox{ iff }
 & \bs{x} \, = \, |\bs{Y_{(N,\varphi,{\vek{z}})}}|
\end{eqnarray*}
is true for all $N\in\NNpos$, for all $\bs{x},\bs{\vek{z}}\in\dom{N}$,
and for the set \,$\bs{Y_{(N,\varphi,\bs{\vek{z}})}} \deff
\setc{\bs{y}\in\dom{N}}{\struc{\dom{N},+,\times}\models \varphi(\bs{y},\bs{\vek{z}})}$.
\end{question_mit}%
Note that the non-counting capability formulated in
Theorem~\ref{theorem:non-counting-ability} does \emph{not} imply a
negative answer to the above question: 
In the highly involved proofs of \cite{FKPS,DeGuSh} it is
essentially used that there are lots of different possible
interpretations of the set $Y$, whereas in
Question~\ref{question:counting_plus_mal} the set $Y$ is defined by a
$\FO(+,\times)$-formula and has thus exactly \emph{one}
interpretation.
\raus{
However, the result of Theorem~\ref{theorem:polylog-counting-ability}
directly implies that the question can be answered ``yes'' under the
additional constraint that $x\leq (\lg N)^c$ for a fixed $c\in\NNpos$.
The according (weaker) version of
Theorem~\ref{theorem:polylog-counting-ability}, formulated for
rudimentary relations, was obtained independently of
\cite{FKPS,DeGuSh} by \cite{ParisWilkie}.
}
\par
In fact, in \cite{ParisWilkie} it was shown that the following \emph{approximate}
counting is indeed possible for rudimentary relations:
For every $\varepsilon>0$ and every $\FO(+,\times)$-formula
$\varphi(y,\vek{z})$ there is a $\FO(+,\times)$-formula
$\chi(x,\vek{z})$ such that the following is true for every
$N\in\NNpos$ and all $\bs{\vek{z}}\in\dom{N}$:
\begin{enumerate}[(1.)\ ]
 \item there is exactly one $\bs{x}\in\dom{N}$ with
   \,$\struc{\dom{N},+,\times}\models\chi(\bs{x},\bs{\vek{z}})$,\quad and
 \item for this $\bs{x}$ we have 
   \ $|\bs{Y_{(N,\varphi,{\vek{z}})}}|\, \leq \,\bs{x}\, <
   \,|\bs{Y_{(N,\varphi,{\vek{z}})}}|^{1+\varepsilon}$. 
\end{enumerate}
\par
Paris and Wilkie \index{Paris, Jeff}\index{Wilkie, Alex} conjecture that 
Question~\ref{question:counting_plus_mal} has a negative answer
(without giving any evidence, except for the fact
that known techniques do not enable us to give a positive answer).
Let us remark, however, that a negative answer would
have the serious complexity theoretic consequence that 
\,$\LinH \neq \ETIME$, \,where $\ETIME$ denotes the class of all problems solvable on
a deterministic Turing machine in linear exponential time
$2^{\bigO(n)}$.
This can be seen as follows: A negative answer to
Question~\ref{question:counting_plus_mal} would imply that
$\FO(+,\times)$ is strictly less expressive than \emph{least fixed
point logic} $\LFP(+,\times)$ on initial segments of $\NN$. However,
it has been mentioned in \cite{AtseriasKolaitis} and proved in
\cite[Theorem~14]{Atserias} that \,$\FO(+,\times)\neq\LFP(+,\times)$
on initial segments of $\NN$ \,if and only if \,$\LinH\neq\ETIME$.
The efforts to separate $\FO$ from $\LFP$ on various kinds of ordered structures are
subsumed under the keyword \emph{the Ordered Conjecture}. \index{Ordered Conjecture}
An overview of what is known
about this conjecture can be found in \cite{AtseriasKolaitis}.
%
%
%
\\
\parno
In the subsequent sections of this paper we consider the expressive power of the logic one
obtains by extending first-order logic with the ability to count.
In Section~\ref{section:PresburgerArithmetic} we will give a positive answer
to the analogue of Question~\ref{question:counting_plus_mal} which
speaks about $\FO(+)$ rather than $\FO(+,\times)$.




\section{First-Order Logic with Counting Quantifiers}\label{section:FO_Count}
In this section we fix the syntax and semantics of first-order logic with counting
quantifiers, and we summarize some important properties of this logic. In particular,
we show that on Skolem arithmetic $\struc{\NN,\times}$ and its initial segments it 
fails to have the isomorphism property.
\subsection{Syntax and Semantics}
\label{section:Presb_and_Counting:syntax}
First-order logic with \emph{unary} counting quantifiers, $\FOunC$, \index{FOunC@$\FOunC$}
is the extension of 
first-order logic obtained by adding unary counting quantifiers of the form
$\exists^{=x}y$. 
For an interpretation $\bs{x}$ of the variable $x$, a formula
$\exists^{=x}y\;\varphi(y)$ expresses that there are exactly $\bs{x}$
many different interpretations $\bs{y}$ of the variable $y$ such that
the formula $\varphi(y)$ is satisfied.
\\
Accordingly, for $k\in\NNpos$, 
first-order logic with $k$-ary counting
quantifiers, $\FOkaryC$, \index{FOkaryC@$\FOkaryC$}
is the extension of first-order logic obtained by adding $k$-ary counting quantifiers
of the form $\exists^{=x_1,\twodots,x_k}y_1,\twodots,y_k$, which allow to count the
number of interpretations of $k$-tuples $(y_1,\twodots,y_k)$ of variables.
\\
To be precise:
Let $k\in\NNpos$, and let $\tau$ be a signature. 
The class of $\FOkaryC(\tau)$-formulas is obtained by the extension of
the calculus for $\FO(\tau)$ via the following rule:
\begin{quote}\em
 If $\varphi$ is a $\FOkaryC(\tau)$-formula and $x_1,\twodots,x_k$ and
 $y_1,\twodots,y_k$ are distinct variables, then
 \,$\exists^{=x_1,\twodots,x_k}y_1,\twodots,y_k\;\varphi$\, is a $\FOkaryC(\tau)$-formula.
\end{quote}
The variables $y_1,\twodots,y_k$ are bounded by this quantifier, whereas the variables 
$x_1,\twodots,x_k$ remain free, i.e.,
$\free(\exists^{=\vek{x}}\vek{y}\;\varphi) = \set{\vek{x}} \cup (\free(\varphi)\setminus\set{\vek{y}})$.
\par
We will evaluate $\FOkaryC(\tau)$-formulas only in structures whose universe is $\ZZ$, $\NN$,
or some initial segment of $\NN$. For such a structure $\A$, the semantics of a
$\FOkaryC(\tau)$-formula of the form $\exists^{=\vek{x}}\vek{y}\;\varphi(\vek{x},\vek{y},\vek{z})$ is defined 
as follows: For interpretations $\bs{\vek{x}}, \bs{\vek{z}} \in A$ of the variables $\vek{x},\vek{z}$
we have
\[
 \begin{array}{c}
   \struc{\A,\bs{\vek{x}},\bs{\vek{z}}}\; \models \;\exists^{=\vek{x}}\vek{y}\;\varphi(\vek{x},\vek{y},\vek{z})
 \medskip\\
   \quad\mbox{iff}\quad
 \medskip\\
   \sum_{i=1}^{k}\bs{x_i}{\cdot} |A|^{k-i} \ = \  
   |\,\setc{(\bs{\vek{y}})\in
   A^k}{\struc{\A,\bs{\vek{x}},\bs{\vek{y}},\bs{\vek{z}}}\models
   \varphi(\vek{x},\vek{y},\vek{z})}\,|\,.
 \end{array}
\]
For \emph{infinite} $A$ this in particular implies that $x_k$ is the only variable in 
$\vek{x}$ which may
be interpreted by a number different from $0$. For finite $A=\dom{N}$,
the formula \,$\exists^{=\vek{x}}\vek{y}\,\varphi$\, expresses that
that the $k$-tuple ${\vek{x}}$ is the $(N{+}1)$-ary
representation of the number of $k$-tuples ${\vek{y}}$ which satisfy $\varphi$.
\\
To denote first-order logic with unary and binary counting
quantifiers, respectively, we write $\FOunC$ and $\FObinC$ \index{FObinC@$\FObinC$}
instead of $\FOonearyC$ and $\FOtwoaryC$.
\subsection{The Isomorphism Property}
\label{section:Presb_and_Counting:soundness}
\index{isomorphism property}
%
For any reasonable logical system one requires it to have the
\emph{isomorphism porperty}. \index{isomorphism property} 
In the present setting this means that the evaluation of a
$\FOkaryC(\tau)$-formula $\varphi(\vek{x})$ makes sense only for $\tau$-structures $\A$
with universe $\ZZ$, $\NN$, or $\dom{N}$ (for some $N\in\NN$), that
have the following property
\begin{center}\em
\begin{tabular}{cl}
 $(*):$ &  If $\pi$ is an automorphism of $\A$ and $\vek{a} \in A$ is an 
 interpretation \\ & of the variables $\vek{x}$, then
 \ $\A\models\varphi(\vek{a})$\; iff 
   \;$\A\models\varphi\big(\pi(\vek{a})\big)$.
\end{tabular}
\end{center}
This property is, of course, true for \emph{rigid} \index{rigid structures} structures, i.e., for structures
which have no automorphisms except for the identity function.
In particular, structures with a discrete linear ordering,
such as $\struc{\ZZ,<}$, $\struc{\NN,<}$, $\struc{\dom{N},<}$, and their extensions,
are rigid. Therefore, \emph{it does make sense to study the expressive power of $\FOkaryC$-formulas on 
those structures.}
\\
\parno
But what about \emph{Skolem arithmetic} \index{Skolem arithmetic}
$\struc{\NN,\times}$ and its initial segments 
$\struc{\dom{N},\times}$? 
In Section~\ref{sec:FO_Arithmetic} we have already seen
that these structures are not rigid. There, we have observed that the mapping
$\pi_{p\gdw q}$ (which interchanges the prime numbers $p$ and $q$ and which leaves fixed all other 
prime numbers), is an automorphism of $\struc{\NN,\times}$, and, as soon as $p,q>\frac{N}{2}$,
even an automorphism of $\struc{\dom{N},\times}$.
However, the non-rigidness does not necessarily imply that $\FOkaryC$
does not have the isomorphism property on these structures.
Nevertheless, for any $k\in\NNpos$, $\FOkaryC(\times)$ does indeed
{neither} have the isomorphism property 
on $\struc{\NN,\times}$ {nor}
on the class of initial segments 
of $\struc{\NN,\times}$:
\begin{proposition_ohne}\label{prop:Mult_IsomProperty}\hspace{4cm}
\begin{enumerate}[(a)\ ]
 \item $\FOunC(\times)$ does not have the isomorphism property on $\struc{\NN,\times}$.
 \item $\FOunC(\times)$ does not have the isomorphism property on 
   {\small $\bigsetc{\struc{\dom{N},\times}}{N\in\NNpos}$}.
   \\ \mbox{ }\fertig
\end{enumerate}
\end{proposition_ohne}%
\begin{proof_mit}
\emph{(a):} \ 
The failure of the isomorphism property of $\FOunC(\times)$ on $\struc{\NN,\times}$ is a direct consequence of 
the fact that
$<$ \ is $\FOunC(\times)$-definable on $\NN$.
I.e., there is a $\FOunC(\times)$-formula $\varphi_<(x,y)$ such that
\ ``$\struc{\NN,\times}\models\varphi_<(a,b)$ iff $a{<}b$'' \ is true for all $a,b\in\NN$. 
For the construction of
the formula $\varphi_<(x,y)$ note that $x{<}y$ is true if and only if $x\neq y$ and 
there are a prime number $p$ and 
$p$-powers $u$ and $v$ such that $u=p^x$, $v=p^y$, and $u\mid v$.
Furthermore,
\begin{enumerate}[$\bullet$]
 \item
   ``$u\mid v$'' can be expressed in $\FO(\times)$ via 
   ``$\exists w\; (u{\times}w{=}v)$'',
 \item
   ``$p$ is a prime number'' can be expressed in $\FO(\times)$ via \\
   ``$p{\neq}1 \;\und\;\forall w\; (w\mid p) \impl (w{=}1\oder w{=}p)$'',
 \item
   ``$u$ is a power of the prime number $p$'' can be expressed in $\FO(\times)$ via \\
   ``$p$ is a prime number $\und\ \forall q\; (q\mid u\,\und\,q\mbox{ is a prime number})
     \impl q{=}p$'',
 \item
   ``$u=p^x$'' can be expressed in $\FOunC(\times)$ via \\
   ``$u$ is a power of the prime number $p$ $\und$
     $\exists^{=x}w\; (w{\neq}u \,\und\, w\mid u)$''.
\end{enumerate}
Altogether, this gives us the desired $\FOunC(\times)$-formula $\varphi_<(x,y)$.
\\ 
To see that the isomorphism property $(*)$ is not satisfied,
let $p,q$ be prime numbers with $p<q$, and let $\pi\deff\pi_{p\gdw q}$ be the 
automorphism of $\struc{\NN,\times}$ which interchanges $p$ and $q$.
Clearly, we have \ $\struc{\NN,\times} \models \varphi_<(p,q)$, \; but
\ $\struc{\NN,\times}\not\models \varphi_<\big(\pi(p),\pi(q)\big)$.
\\ \par
\emph{(b):} \ 
Note that the formula $\varphi_<(x,y)$ of part \emph{(a)} is of no use here, because it 
gives us ``$<$'' only for numbers of size up to $\lg N$ when $\dom{N}$ is the underlying universe 
--- and from Lemma~\ref{lemma:kleiner_auf_Anfangsstueck} we know that ``$<$'' is 
$\FO(\times)$-definable even for numbers of size up to $\sqrt{N}$.
However, the failure of the isomorphism property of the logic $\FOunC(\times)$ 
on the class \,$\bigsetc{\struc{\dom{N},\times}}{\allowbreak N\in\NNpos}$
can be obtained as follows:
Consider the $\FOunC(\times)$-formula 
\begin{eqnarray*}
\psi(x) & \deff &
\exists^{=x}y\; \nicht (\mbox{$y$ is a prime number}).
\end{eqnarray*}
Of course we have, for all $N\in\NNpos$ and all
interpretations $a\in\dom{N}$ of the variable $x$, that
\,$   \struc{\dom{N},\times}\models\psi(a)
 \ \mbox{ iff }
 \ a \, = \, |\setc{b\in\dom{N}}{b \mbox{ is not a prime number}}|$.\\
However, for $N\deff 8$ and $p\deff 5$ and $q\deff 7$, the mapping $\pi\deff\pi_{p\gdw q}$ 
is an
automorphism of $\struc{\dom{N},\times}$, for which the property $(*)$ describing 
the isomorphism property is not satisfied: The set of non-prime numbers in 
$\dom{N}$
is $\set{0,1,4,6,8}$. This set has 
cardinality $p{=}5$, and thus 
we have \ $\struc{\dom{N},\times} \models \psi(p)$, \; but
\ $\struc{\dom{N},\times}\not\models \psi\big(\pi_{p\gdw q}(p)\big)$.
\par
Let us mention that from 
the \emph{Prime Number Theorem} (cf., e.g., \cite{Zahlentheorie})
it follows that for any $N_0$ there are
a $N\geq N_0$ and two different prime numbers $p,q$ with $\frac{N}{2}<p,q\leq N$ such
that \ $\struc{\dom{N},\times} \models \psi(p)$, \; but
\ $\struc{\dom{N},\times}\not\models \psi\big(\pi_{p\gdw q}(p)\big)$. 
I.e., the isomorphism property of $\FOunC(\times)$ cannot be obtained by resticting considerations to 
initial segments that are ``large enough''.
\end{proof_mit}%
\subsection{Easy Facts and Known Results}\label{section:Presb_and_Counting:known}
For the rest of this paper we will concentrate on first-order logic
with counting quantifiers on rigid structures such as
$\struc{\dom{N},<}$ and $\struc{\dom{N},+}$.
It is obvious that
\begin{quote}\bf
  $\bs{+}$ is definable in $\bs{\FOunC(<)}$\\ 
  on initial segments of $\NN$, on finite structures, and on $\NN$,
\end{quote} 
via the formula \,$\varphi_+(x,y,z)\deff \exists^{=y}u\,(x<u\leq z)$.
Furthermore, 
\begin{quote}\bf
  $\bs{\times}$ is definable in $\bs{\FObinC(<)}$ \index{FObinC<@$\FObinC(<)$}\\
  on initial segments of $\NN$, on finite structures, and on $\NN$, 
\end{quote} 
via the formula \,$\varphi_{\times}(x,y,z)\deff
\exists^{=0,z}u,v\,(1\leq u\leq x\ \und\ 1\leq v\leq y)$.
This is true because \,
{\small\[
 x\times y\ =\ \sum_{u=1}^{x}y\ =\ \sum_{u=1}^{x}\sum_{v=1}^{y}1\ =\ 
 |\setc{(u,v)}{1\leq u\leq x\ \und\ 1\leq v\leq y}|\,. 
\]}%
It is not difficult to see the following:
\begin{proposition_mit}\label{proposition:counting_NN}\mbox{ }\\
For all $k\in\NNpos$, \ 
{${\FOkaryC(+,\times) = \FO(+,\times)}$ on $\NN$}.
\end{proposition_mit}%
\begin{proof_mit}
We encode a finite set $Y$ by the unique number 
$u$ which satisfies, for all $y\in\NN$, that $\Bit(u,y)$ iff $y\in Y$.
The $\FO(+,\times)$-formula $\varphi_{\Bit}(u,y)$ from Theorem~\ref{theorem:arithmetic} hence
expresses that $y$ belongs to the set encoded by $u$.
Furthermore, from the counting capability of
Theorem~\ref{theorem:polylog-counting-ability} we obtain a $\FO(+,\times)$-formula
$\varphi_{\textsl{BITSUM}}(x,u)$ expressing that $x$ is the
number of $y\in\NN$ which satisfy $\Bit(u,y)$. I.e.,
$\varphi_{\textsl{BITSUM}}(x,u)$ expresses that $x$ is the
cardinality of the set encoded by $u$.
\par
Now, a given $\FOunC(+,\times)$-formula
\,$\exists^{=x}y\,\psi(x,y,\vek{z})$\, is equivalent over $\NN$ to the
$\FO(+,\times)$-formula
\[
   \exists u\ \Big(\varphi_{\textsl{BITSUM}}(x,u)\ \und\ 
   \forall y\,\big( \varphi_{\Bit}(u,y) \gdw \psi(x,y,\vek{z}) \big)\Big)\,.
\]
Here, $u$ encodes the set of all $y$ satisfying $\psi$.
\\
For a given $\FOkaryC(+,\times)$-formula 
it hence suffices to find an equivalent formula in $\FOunC(+,\times)$.
\par
We encode a tuple $(y_1,\twodots,y_k)\in\NN^k$ by the single number 
$v = p_1^{y_1}\cdots p_k^{y_k}$, where $p_i$ denotes the $i$-th
largest prime number (for $i\in\set{1,\twodots,k}$).
A given $\FOkaryC(+,\times)$-formula 
\,$\exists^{=x_1,\twodots,x_k}y_1,\twodots,y_k\,\psi(\vek{x},\vek{y},\vek{z})$\,
is thus equivalent over $\NN$ to an $\FOunC(+,\times)$-formula which
expresses that
\[
   x_1{=}0 \ \und \cdots \und\ x_{k-1}{=}0\ \und\ 
   \exists^{=x_k}v\ \big( \exists y_1 \twocdots \exists y_k\ 
      v=p_1^{y_1}\cdots p_k^{y_k} \ \und\ \psi(\vek{x},\vek{y},\vek{z})\big)\,.
\]
This completes the proof of Proposition~\ref{proposition:counting_NN}.
\end{proof_mit}%
\\
\parno
Note that the above proof does not work for initial segments of $\NN$,
because the number $u$ which encodes a finite set $Y$ is exponentially
larger than the elements of $Y$. 
Indeed, it is still
open whether $\FO(+,\times) = \FOunC(+,\times)$ \emph{on initial segments of
$\NN$}. However, from Theorem~\ref{theorem:non-counting-ability} we
know that $\FO(+,\times) \neq \FOunC(+,\times)$ \emph{on finite structures}.
\\
\parno
It was shown in \cite{BIS} that, for ordered finite structures
over arbitrary signatures $\tau$, the class of problems definable in
$\FOunC(+,\times,\tau)$ is exactly the (logtime-uniform version of the) circuit complexity 
class $\TCzero$.\footnote{By 
definition, the class $\TCzero$\index{TC0@$\TCzero$} 
(in the literature sometimes also denoted
$\class{ThC}^0$)
consists of all problems solvable by uniform polynomial size,
constant depth circuits of AND, OR, NOT, and THRESHOLD gates of
unbounded fan-in.}
It is a deep result, following from \cite{BIS}, that for all $k\in\NNpos$,
\begin{quote}\bf
  $\bs{\FOkaryC(+,\times) = \FOunC(+,\times)}$ \\ 
  on finite structures and on initial segments of $\NN$.
\end{quote} 
Actually, in Proposition~10.3 of \cite{BIS} it is shown that a
\emph{binary counting quantifier} can be expressed using \emph{unary
  majority quantifiers} \index{majority quantifier} and the $\Bit$ predicate. 
Here, a unary majority quantifier \,$My\,\varphi(y)$\, expresses that more than half of the
interpretations of $y$ do satisfy $\varphi(y)$. \index{FOunM@$\FOunM$} 
The proof of \cite{BIS} easily generalizes from \emph{binary} to \emph{$k$-ary}
counting quantifiers, leading to the result that 
\,$\FOkaryC(+,\times) = \FOunC(+,\times) = \FOunM(<,\allowbreak\Bit) = \TCzero$.
(Note that the unary majority quantifier \,$My\,\varphi(y)$\, can
easily be expressed using unary counting via
\,$\exists u\,\exists v\,u{>}v\,\und\,\exists^{=u}y\,\varphi(y)\,\und\,
   \exists^{=v}y\,\nicht\varphi(y)$.)
\par
Barrington, \index{Barrington, David A.\ Mix} 
Immerman, \index{Immerman, Neil}
and Straubing \index{Straubing, Howard}
\cite{BIS} 
also gave a logical characterization of the class $\TCzero$
which does not need the $\Bit$ predicate, i.e., which does not need
$+$ and $\times$:
They proved that $\TCzero = \FObinM(<)$ on finite structures. Here, 
$\FObinM$ \index{FObinM@$\FObinM$}
is the extension of
first-order logic obtained by adding \emph{binary majority quantifiers} of
the form $M x,y\,\varphi(x,y)$, expressing that more than half of the
interpretations of $(x,y)$ do satisfy $\varphi(x,y)$.
\par
In \cite[Corollary\,4.4]{LMSV} it was shown that 
$\FOunM(<) \subsetneq \FObinM(<)$ on finite structures.
Although formulated in the terminology of certain 
\emph{groupoidal Lindstr\"{o}m quantifiers}, \index{Lindstr\"{o}m quantifier} 
their proof basically shows the following:
For \emph{pure} arithmetic on initial segments of $\NN$, all $\FOunM(<)$-definable spectra
are also definable in $\FO(<,+)$.
\par
Concerning the power of $\FO(<,+)$ for \emph{pure} arithmetic, the 
main result of the following section goes one step further:
In Theorem~\ref{maintheorem:Presb_Unary_Counting},
Corollary~\ref{corollary:Presb_Unary_Counting_NN}, and
Corollary~\ref{corollary:Presb_Unary_Counting_finite} we will show that
\begin{quote}\bf
$\bs{\FO(<,+) \,=\, \FOunC(<,+)}$ \index{FOunC<+@$\FOunC(<,+)$}
\\ on $\ZZ$, on $\NN$, and on initial segments of $\NN$.
\end{quote}
Altogether, we now have a complete picture of the
expressive power of first-order logic with counting quantifiers and
arithmetic. This picture is
visualized in Figure~\ref{figure:Inclusion_Diagram_pure} and
Figure~\ref{figure:Inclusion_Diagram_finite_structures} at the end of this paper.




\section{Presburger Arithmetic is Closed Under Unary Counting Quantifiers}\label{section:PresburgerArithmetic}
In this section we show that ${\FOunC(<,+) = \FO(<,+)}$ on initial segments of $\NN$, 
on $\NN$, and on $\ZZ$.
An important tool for our proof 
will be \emph{Presburger's quantifier elimination} \cite{Presburger} \index{Presburger@Presburger, M.}
\index{quantifier elimination}\index{Presburger's quantifier elimination}
which states the following:
\\
Every $\FO(<,+)$-formula $\varphi(\vek{x})$ is equivalent over $\ZZ$ to
a Boolean combination of atoms of the form $t=t'$, $t< t'$, and
$t\equiv_n t'$, where\footnote{Recall that
$\equiv_n$ denotes the congruence relation modulo $n$.} $t$ and $t'$ are terms built from 
the constants $0$ and $1$, 
the variables $\vek{x}$, and the addition \emph{function} $f_+$.
Essentially this means that $\FO(<,+)$ over $\ZZ$
can express equality, inequality, and residue classes of
terms --- and nothing else!
A well-presented proof of Presburger's quantifier elimination can be found, e.g., in the 
textbook \cite[Chapter~III.4]{Smorynski}.
\subsection{Basic Facts Concerning Presburger Arithmetic}
\index{Presburger arithmetic}
We define the \emph{Presburger signature} $\Presb$ \index{Presb@$\Presb$} 
to consist of all predicates 
needed for Presburger's quantifier elimination. I.e., 
\ \( 
   \Presb  \deff  \set{\,0,\,1,\,f_+,\,<,\,({\equiv_n})_{n\in\NNpos}\;}
\) \
consists of constant symbols $0$ and
$1$, a binary function symbol $f_+$, a binary relation symbol $<$, and binary relation 
symbols $\equiv_n$, for every $n\in\NNpos$.
When considered over the universe $\ZZ$ or $\NN$, these 
symbols are always interpreted in the natural way via the numbers $0$ and $1$, the 
addition function, the linear ordering, and the congruence relation modulo $n$. 
It should be obvious that these predicates are 
$\FO(+)$-definable in $\struc{\NN,+}$ and
$\FO(<,+)$-definable\footnote{Note that $+$ alone is not sufficient
  here, because the order relation ``$<$'' (respectively, the unary
  relation ``${>}0$'') are not $\FO(+)$-definable in $\struc{\ZZ,+}$.} in $\struc{\ZZ,<,+}$.
Speaking about \emph{Presburger arithmetic}, we therefore refer to one of the structures 
$\struc{\ZZ,\Presb}$,\,
$\struc{\ZZ,<,+}$,\, $\struc{\NN,\Presb}$,\, $\struc{\NN,+}$.
\\
From Presburger's quantifier elimination we
know that the structure $\struc{\ZZ,\Presb}$ has quantifier elimination. I.e.,
every $\FO(\Presb)$-formula is equivalent over $\struc{\ZZ,\Presb}$ to a
Boolean combination of atomic $\Presb$-formulas.
Moreover, in this Boolean combination of atoms, the negation $\nicht$ is not needed,
because
\begin{itemize}
 \item
   $\nicht t_1{=}t_2$\quad can be replaced by  
   \quad $t_1{<}t_2 \,\oder\, t_2{<}t_1$\,,
 \item
   $\nicht t_1{<}t_2$\quad can be replaced by  
   \quad $t_1{=}t_2 \,\oder\, t_2{<}t_1$\,, \quad and
 \item
   $\nicht t_1 {\equiv_n} t_2$\quad can be replaced by \\
   $t_1 \equiv_n t_2{+}1 \ \oder\ 
      t_1 \equiv_n t_2{+}1{+}1 \ \oder\cdots\oder\ 
      t_1 \equiv_n t_2{+}(n{-}1){\cdot}1$\,.
\end{itemize}
Hence Presburger's quantifier elimination can be formulated as follows:
\index{quantifier elimination}\index{Presburger's quantifier elimination}
\begin{theorem_mit}[Presburger's Quantifier
  Elimination]\label{theorem:Presburger}\quad
Every $\FO(\Presb)$-formula $\varphi(\vek{z})$ is equivalent over
$\struc{\ZZ,\Presb}$ to a formula of the form
\ \(
   \Oder_{i=1}^{m}\Und_{j=1}^{n_i} \alpha_{i,j}(\vek{z})\,,
\) \ 
where the $\alpha_{i,j}$ are atoms built from the symbols in
$\set{=}\cup\Presb\cup\set{\vek{z}}$. 
\end{theorem_mit}%
%
%
In order to gain full understanding of Presburger arithmetic, let us
have a look at what the $\Presb$-atoms may express:
\par
Let $y$ and $\vek{z}= z_1,\twodots,z_\nu$ be distinct first-order
variables. 
A $\Presb$-atom $\alpha(y,\vek{z})$ is built from the symbols in
$\set{=}\cup\setc{0,1,f_+,<,\equiv_n}{n\in\NNpos}\cup\set{y,\vek{z}}$.
For better readability we will write $+$ instead of $f_+$. I.e.,
$\alpha$ is of the form
\begin{eqnarray*}
  (*):\qquad\qquad u_1 + \cdots + u_k & \ltimes & v_1 + \cdots + v_l
\end{eqnarray*}
where ${\ltimes}$ is an element in $\setc{<,=,\equiv_n}{n\in\NNpos}$, \,and
$u_1,\twodots,u_k,v_1,\twodots,v_l$ are (not necessarily distinct)
elements in $\set{0,1,y,\vek{z}}$.\\
Let $m_1,m_y,m_{z_1},\twodots,m_{z_\nu}$ be the number of occurrences
of the constant $1$, the variable $y$, and the variables
$z_1,\twodots,z_\nu$, respectively, on the left side of $(*)$. Similarly, let
$n_1,n_y,\allowbreak n_{z_1},\twodots,\allowbreak n_{z_\nu}$ be the corresponding
multiplicities for the right side of $(*)$. Interpreted in the structure
$\struc{\ZZ,\Presb}$, the atom $(*)$ expresses that
\begin{eqnarray*}
   m_1{\cdot} 1\ + \ m_y{\cdot} y\ + \ \sum_{j=1}^{\nu} m_{z_j}{\cdot} z_j
 & \ \ \ltimes\ \ 
 & n_1{\cdot} 1\ + \ n_y{\cdot} y\ + \ \sum_{j=1}^{\nu} n_{z_j}{\cdot} z_j\,,
\end{eqnarray*}
which is equivalent to
\ \(
(m_y{-}n_y){\cdot} y
\ \ltimes\ 
(n_1{-}m_1){\cdot} 1\ + \ \sum_{j=1}^{\nu}(n_{z_j}{-}m_{z_j}){\cdot} z_j\,.
\)
\\
I.e., there are $c,d,k_1,\twodots,k_{\nu}\in\ZZ$ such that $(*)$ is equivalent to 
\begin{eqnarray*}
   (**):\qquad
   c\cdot y & \ltimes & d + \sum_{j=1}^{\nu}k_j z_j\,.
\end{eqnarray*}
In case $c=0$, $(**)$ is equivalent to \ $0\ltimes d+\sum_{j=1}^{\nu}k_jz_j$.
In case $\ltimes\in\set{<,=}$, $(**)$ is equivalent to 
\ $y\ltimes \frac{1}{c} \big( d+\sum_{j=1}^{\nu}k_jz_j \big)$ \ if $c>0$, \ and to
\ $\frac{1}{c} \big( d+\sum_{j=1}^{\nu}k_jz_j \big) \ltimes y$ \ if $c<0$.
\raus{
\[
\begin{array}{rll}
  \null\quad\bullet\ 
& 
  0 \quad\ltimes\quad
  (n_1{-}m_1){\cdot} 1\ + \ \sum_{j=1}^{\nu}
  (n_{z_j}{-}m_{z_j}){\cdot} z_j 
& \quad \mbox{if \ } n_y=m_y\,,
\\[5ex]
  \null\quad\bullet\ 
& \displaystyle
  y \quad\ltimes\quad
  \frac{
    (n_1{-}m_1){\cdot} 1\ + \ \sum_{j=1}^{\nu}
    (n_{z_j}{-}m_{z_j}){\cdot} z_j
  }{
    m_y-n_y}
& \quad \mbox{if \ } n_y<m_y\,,
\\[5ex]
  \null\quad\bullet\ 
& \displaystyle   
  \frac{
    (n_1{-}m_1){\cdot} 1\ + \ \sum_{j=1}^{\nu}
    (n_{z_j}{-}m_{z_j}){\cdot} z_j
  }{
    m_y-n_y}
  \quad\ltimes\quad y
& \quad \mbox{if \ } n_y>m_y\,.
\end{array}
\]
}
\\
It remains to consider the case where $c\neq 0$ and $\ltimes$ is a congruence relation 
$\equiv_n$ for some $n\in\NNpos$.
The following Lemma~\ref{lemma:equiv-Presb} shows that in this case there are 
$d',k'_1,\twodots, k'_{\nu}\in\ZZ$ and $c',n'\in\NNpos$ such that $(**)$ is 
equivalent to \ $y \equiv_{n'} \frac{1}{c'}\big( d' + \sum_{j=1}^{\nu}k'_j z_j\big)$.
\begin{lemma_mit}\label{lemma:equiv-Presb} 
Let $0\neq c\in\ZZ$ and $n\in\NNpos$. Let $g$ be the greatest common divisor of 
$c$ and $n$, and let $c'\deff\frac{c}{g}$ and $n'\deff\frac{n}{g}$. 
Since $c'$ and $n'$ are relatively prime there must exist a $c''\in\NNpos$ such that
\ $c' c'' \equiv_{n'} 1$.
\\
The following is true for all $y,e\in\ZZ$:\quad
$c y \equiv_n e$\quad iff\quad $y \equiv_{n'} c'' \frac{e}{g}$.
\end{lemma_mit}%
\begin{proof_mit}
Clearly, $y\equiv_{n'} c''\frac{e}{g}$ \ iff \ $c'y \equiv_{n'} c'c''\frac{e}{g}$ which, since
$c'c''\equiv_{n'}1$, is equivalent to $c'y\equiv_{n'}\frac{e}{g}$.
Furthermore, $c'y\equiv_{n'} \frac{e}{g}$ \ iff \ there is a $k\in\ZZ$ such that 
$c'y=n'k+\frac{e}{g}$ \ iff \ $gc'y = gn'k+e$ \ iff \ $cy= nk+e$ \ iff \ $cy\equiv_n e$. 
\end{proof_mit}%
\medskip\\
To denote a fraction of the form
\ $\frac{1}{c} \big(d+\sum_{j=1}^{\nu}k_j z_j \big)$ \ with $c,d,k_1,\twodots,k_{\nu}\in\ZZ$ and $c\neq 0$,
we will write $t(\vek{z})$ for short,
and we will call such fractions 
\emph{generalized $\Presb$-terms} \index{generalized $\Presb$-term} 
over the variables $\vek{z}$.
What we have just seen above is the following:
\begin{fact_mit}[$\Presb$-Atoms]\label{fact:Presb_Atoms_Normal_Form}\rm
Let $y$ and $\vek{z}= z_1,\twodots,z_\nu$ be distinct first-order
variables. For every $\Presb$-atom $\alpha(y,\vek{z})$ there is a
generalized $\Presb$-term $t(\vek{z})$ or a $\Presb$-atom
$\beta(\vek{z})$, in which the variable $y$ does not occur, such that
$\alpha(y,\vek{z})$ expresses over $\struc{\ZZ,\Presb}$ that 
\medskip\\
\begin{tabular}{rll}
   \quad $\bullet$
 & $y>t(\vek{z})$ 
 & \ (\emph{lower bound} on $y$),
 \medskip\\
   \quad $\bullet$
 & $y<t(\vek{z})$ 
 & \ (\emph{upper bound} on $y$),
 \medskip\\
   \quad $\bullet$
 & $y \equiv_n t(\vek{z})$ 
 & \ (\emph{residue class} of $y$, for an
   appropriate $n\in\NNpos$),
 \medskip\\
   \qquad $\bullet$
 & $y= t(\vek{z})$ 
 & \ (\emph{equation} for $y$),\quad or
 \medskip\\
   \quad $\bullet$
 & $\beta(\vek{z})$ 
 & \ (\emph{independent} of $y$).
\end{tabular}
\medskip\\
On the other hand, it is straightforward to see that for any  
$\ltimes\in\setc{>,<,=,\equiv_n}{n\in\NNpos}$ and
any generalized $\Presb$-term $t(\vek{z})$, the generalized atom
$y \ltimes t(\vek{z})$ can be expressed by a quantifier free
$\FO(\Presb)$-formula. Similarly, for $\ltimes\in\set{>,<,=}$, also the
generalized atoms $y \ltimes \aufgerundet{t(\vek{z})}$ and $y \ltimes
\abgerundet{t(\vek{z})}$ can be expressed by quantifier free $\FO(\Presb)$-formulas.
\end{fact_mit}%
\subsection{$\struc{\ZZ,\Presb}$ and Unary Counting Quantifiers}
In this section we prove that
Presburger's quantifier elimination can be extended to unary counting quantifiers:
\begin{theorem_mit}[Elimination of Unary Counting
  Quantifiers]\label{maintheorem:Presb_Unary_Counting} \mbox{}\\
Every $\FOunC(\Presb)$-formula $\varphi(\vek{z})$ is equivalent over
$\struc{\ZZ,\Presb}$ to a formula of the form
\ \(
   \Oder_{i=1}^{m}\Und_{j=1}^{n_i} \alpha_{i,j}(\vek{z})\,,
\) \ 
where the $\alpha_{i,j}$ are atoms built from the symbols in \linebreak[4]
$\set{=}\cup\Presb\cup\set{\vek{z}}$. 
\end{theorem_mit}%
In particular, this means that \
{\bf
$\bs{\FOunC(<,+)\ =\ \FO(<,+)}$ \ on $\bs{\ZZ}$.} 
%
\\
\parno
The proof of Theorem~\ref{maintheorem:Presb_Unary_Counting} will
be given in a series of lemmas, the first (and most laborious to
prove) is the following:
\begin{lemma_mit}\label{lemma:Presb_Konjunktionen}
Every $\FOunC(\Presb)$-formula of the form
\ \(
   \exists^{=x}y\; \Und_{j=1}^{n} \alpha_j(y,\vek{z})\,,
\) \ 
where the $\alpha_j$ are atoms built from the symbols in
$\set{=}\cup\Presb\cup\set{y,\vek{z}}$, 
is equivalent over $\struc{\ZZ,\Presb}$ to a $\FO(\Presb)$-formula.
\end{lemma_mit}%
Before proving Lemma~\ref{lemma:Presb_Konjunktionen} let us first look
at an example that exposes all the relevant proof ideas.
\begin{example_mit}\label{example:Presb_Konjunktionen}\rm
Consider the formula \,$\varphi(x,\vek{z})\ \deff$
\[
  \exists^{=x}y\;
  (z_2{+}y<z_3{+}y{+}y)\,\und\,
  (y<z_2)\,\und\,
  (y{+}y{+}y<z_1)\,\und\,
  (y{+}z_4 \equiv_4 z_3)\,.
\]
For interpretations $\bs{x,\vek{z}}$ in $\ZZ$ of the variables
$x,\vek{z}$, this formula expresses that there are exactly $\bs{x}$
many different $\bs{y}\in\ZZ$ which satisfy the constraints
\[ 
  (*):\qquad
  \underbrace{\bs{y}\, >
  \,\bs{z_2}{-}\bs{z_3}\vphantom{{\frac{\bs{z_1}}{3}}}}_{\textrm{lower bound}} 
  \,,\quad
  \underbrace{\bs{y}\, < \,\bs{z_2}\,,\quad
              \bs{y}\, < \,{\frac{\bs{z_1}}{3}}}_{\textrm{upper bound}}
  \,,\quad\mbox{and}\quad
  \underbrace{\bs{y}\, \equiv_4 \,\bs{z_3}{-}\bs{z_4}\vphantom{\aufgerundet{\frac{\bs{z_1}}{3}}}}_{\textrm{residue class}}.
\]
We consider the integers $\textbf{\itshape low}\deff \bs{z_2}{-}\bs{z_3}$,\ 
$\textbf{\itshape up}\deff\min\set{\bs{z_2},\;
  \aufgerundet{\frac{\bs{z_1}}{3}}}$,\ and
$\textbf{\itshape first}$, where $\textbf{\itshape first}$ is the smallest integer $>$
$\textbf{\itshape low}$ which belongs to the correct residue class, i.e.\ which
satisfies \,$\textbf{\itshape first}\,\equiv_4 \,\bs{z_3}{-}\bs{z_4}$. 
The constraints $(*)$ can be visualized as shown in
Figure~\ref{figure:example:Presb_Konjunktionen}.
\begin{figure}[!htbp]
\bigskip
\begin{center}
\fbox{
\scalebox{0.65}{
\begin{pspicture}(-1.75,-1.75)(15.75,+1.25)
\rput(2.5,0){\rnode{l}{\huge{(}}}%
\rput(2.5,-0.75){\rnode{low}{{\textbf{\itshape low}}}}%
\rput(4,0){\rnode{f}{\huge{[}}}%
\rput(4,-0.8){\rnode{first}{{\textbf{\itshape first}}}}%
\rput(13,0){\rnode{u}{\huge{)}}}%
\rput(13,-0.75){\rnode{up}{{\textbf{\itshape up}}}}%
\pnode(-1,0){Links}%
\pnode(15,0){Rechts}%
\ncline{-}{Links}{Rechts}%
\psdot[fillcolor=black,linecolor=black,dotsize=6pt](0,0)%
\psdot[fillcolor=black,linecolor=black,dotsize=6pt](2,0)%
\psdot[fillcolor=black,linecolor=black,dotsize=6pt](4,0)%
\psdot[fillcolor=black,linecolor=black,dotsize=6pt](6,0)%
\psdot[fillcolor=black,linecolor=black,dotsize=6pt](8,0)%
\psdot[fillcolor=black,linecolor=black,dotsize=6pt](10,0)%
\psdot[fillcolor=black,linecolor=black,dotsize=6pt](12,0)%
\psdot[fillcolor=black,linecolor=black,dotsize=6pt](14,0)%
\end{pspicture}
}
}
\caption{\small Visualization of the constraints $(*)$. The black
points are those which belong to the correct residue class; the black
points in the interval $[{\textbf{\itshape
    first},\textbf{\itshape up}})$ are exactly those integers $\bs{y}$
which satisfy the constraints $(*)$.}\label{figure:example:Presb_Konjunktionen}
\end{center}
\end{figure}
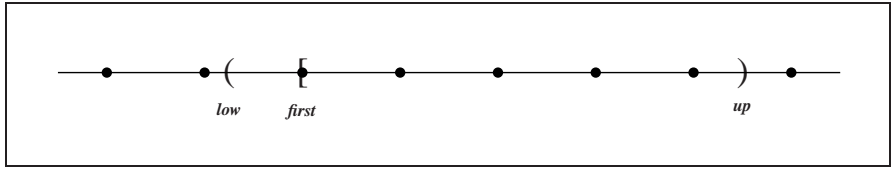
\\
From Figure~\ref{figure:example:Presb_Konjunktionen} one can directly
see that there are exactly 
\ \( 
  \max \left\{\,0\; ,\; 
  \aufgerundet{\frac{\textbf{\itshape up}-\textbf{\itshape first}}{4}}
  \right\}
\) \ 
many different $\bs{y}\in\ZZ$ which satisfy the constraints $(*)$.
Hence, the statement
\emph{``there are exactly $\bs{x}$ many $\bs{y}\in\ZZ$ which
 satisfy the constraints $(*)$''}
can be expressed by the $\FO(\Presb)$-formula \,$\psi(x,\vek{z})\ \deff$
\[ 
 \begin{array}{l}
   \exists\; \textit{low}\ \exists\; \textit{up}\ \exists\;
   \textit{first}\ 
 \\[1ex] 
   \qquad
   \big(\textit{low} = z_2{-}z_3\big)\ \ \und
 \\[1ex] 
   \qquad
   \big(\textit{up}=z_2\,\oder\,\textit{up}= \aufgerundet{\frac{z_1}{3}} \big)
   \ \ \und\ \ 
   \big(\textit{up}\leq z_2\big)\ \ \und\ \ \big(\textit{up}\leq
   \aufgerundet{\frac{z_1}{3}}\big)\ \ \und
 \\[1ex]
   \qquad
   \big(\textit{first}>\textit{low}\big)\ \ \und\ \ 
   \big(\textit{first} \equiv_4 z_3{-}z_4\big)\ \ \und\ \ 
 \\[1ex]
   \qquad
   \big(\forall v\ 
   (v>\textit{low}\,\und\, v \equiv_4 z_3{-}z_4) \,\impl\, v\geq
   \textit{first}\,\big)\ \ \und\ \ 
 \\[1ex]
   \qquad
   \big(\textit{up}\leq\textit{first} \,\impl\, x{=}0\big)\ \ \und\ \  
   \big(\textit{up}>\textit{first} \,\impl\, x= \aufgerundet{\frac{\textit{up}-\textit{first}}{4}}\,\big)
   \,.
 \end{array}
\]
Altogether, we have constructed a $\FO(\Presb)$-formula
$\psi(x,\vek{z})$ which is equivalent over $\struc{\ZZ,\Presb}$
to the $\FOunC(\Presb)$-formula $\varphi(x,\vek{z})$.
\end{example_mit}%
Using the ideas presented in
Example~\ref{example:Presb_Konjunktionen}, we are now ready for
the formal proof of Lemma~\ref{lemma:Presb_Konjunktionen}.
\\
\parno
\begin{proofc_mit}{of Lemma~\ref{lemma:Presb_Konjunktionen}}
\mbox{ }\\
Let \,$\varphi(x,\vek{z})\ \deff\
\exists^{=x}y\,\Und_{j=1}^{n}\alpha_j(y,\vek{z})$\,
be the given formula, where the $\alpha_j$ are atoms built from
the symbols in $\set{=}\cup\Presb\cup\set{y,\vek{z}}$.
Our aim is to construct a $\FO(\Presb)$-formula $\psi(x,\vek{z})$
which is equivalent to $\varphi(x,\vek{z})$ over $\struc{\ZZ,\Presb}$.
\\
The atoms $\alpha_1,\twodots,\alpha_n$ impose constraints on $y$.
According to Fact~\ref{fact:Presb_Atoms_Normal_Form} we can partition
the set of atoms $\set{\alpha_1,\twodots,\alpha_n}$ into 
\begin{itemize}
 \item 
   a set $L$ consisting of all atoms $\alpha_j$ which express a \emph{lower
   bound} of the form $y > t_j(\vek{z})$,
 \item 
   a set $U$ consisting of all atoms $\alpha_j$ which express an \emph{upper
   bound} of the form $y < t_j(\vek{z})$,
 \item 
   a set $R$ consisting of all atoms $\alpha_j$ which express a \emph{residue
   class} of the form $y \equiv_{n_j} t_j(\vek{z})$,
 \item 
   a set $E$ consisting of all atoms $\alpha_j$ which express an \emph{equation}
   of the form $y = t_j(\vek{z})$,
 \item 
   a set $I$ consisting of all atoms $\alpha_j$ which are \emph{independent}
   of $y$, i.e.\ which are equivalent over $\struc{\ZZ,\Presb}$
   to an atom $\beta_j(\vek{z})$ in which the variable $y$ does not occur.
\end{itemize}
For interpretations $\bs{x,\vek{z}}$ in $\ZZ$ of the variables
$x,\vek{z}$, the formula 
\[ 
\varphi(x,\vek{z})\ \deff\  
\exists^{=x}y\;\Und_{j=1}^{n}\alpha_j(y,\vek{z})
\] 
expresses that
there are exactly $\bs{x}$ many different $\bs{y}\in\ZZ$ which satisfy
all the constraints $(*)$ in $L$, $U$, $R$, $E$, and $I$.
\\
\parno
We first consider the easy case where $E\neq \emptyset$.\\
W.l.o.g.\ $\alpha_1\in E$. This means that an $\bs{y}\in\ZZ$ which satisfies the
constraints $(*)$ must in particular satisfy the constraint
$y=t_1(\vek{z})$.
Hence there is at most one $\bs{y}\in\ZZ$ (namely, $\bs{y}\deff
t_1(\bs{\vek{z}})$) that satisfies all the constraints
$(*)$; and the formula \,$\varphi(x,\vek{z})\, \deff\,
\exists^{=x}y\,\Und_{j=1}^{n}\alpha_j(y,\vek{z})$\, is equivalent to the
$\FO(\Presb)$-formula \,$\psi(x,\vek{z})\deff$
\[
\begin{array}{l}
  \big(x=0 \ \oder\ x=1\big)\ \ \und\ \  \Big( x=1 \ \gdw\ 
\\[2ex]
  \qquad
  \displaystyle
  \Und_{j:\alpha_j\in L} \ \ t_1(\vek{z})>t_j(\vek{z})\ \ \und\ 
  \Und_{j:\alpha_j\in U} t_1(\vek{z})<t_j(\vek{z})\ \ \und\ 
\\[2ex]
  \qquad
  \displaystyle
  \Und_{j:\alpha_j\in R}\! t_1(\vek{z})\equiv_{n_j} t_j(\vek{z})\,\,\ \und\ 
  \Und_{j:\alpha_j\in E} t_1(\vek{z})=t_j(\vek{z})\ \ \und\ 
  \Und_{j:\alpha_j\in I} \beta_j(\vek{z})\,\Big)\,.
\medskip
\\
\end{array}
\]
Let us now consider the case where $E=\emptyset$.\\
First of all, we simplify the constraints in $L$, $U$, and $R$:
If $L\neq \emptyset$, the constraints in $L$ can be replaced by the
single constraint $y>\textit{low}$, where the variable $\textit{low}$
is enforced to be interpreted by the maximum lower bound on $y$ via
the $\FO(\Presb)$-formula
\begin{eqnarray*}
  \psi_{\textit{low}}(\textit{low},\vek{z})
& \,\deff\,
& \Big( \Oder_{j:\alpha_j\in L} \!\!\!\!\textit{low} =
  \abgerundet{t_j(\vek{z})}\,\Big)\ \und\ 
  \Und_{j:\alpha_j \in L}\!\! \textit{low} \geq
  \abgerundet{t_j(\vek{z})}\,.
\end{eqnarray*}
Similarly, if $U\neq\emptyset$, we can replace the constraints in $U$
with the single constraint $y<\textit{up}$, where the variable
$\textit{up}$ is enforced to be interpreted by the minimum upper bound on
$y$ via the $\FO(\Presb)$-formula
\begin{eqnarray*}
  \psi_{\textit{up}}(\textit{up},\vek{z})
& \,\deff\,
& \Big( \Oder_{j:\alpha_j\in U} \!\!\textit{up} =
  \aufgerundet{t_j(\vek{z})}\,\Big)\ \und\ 
  \Und_{j:\alpha_j \in U} \textit{up} \leq \aufgerundet{t_j(\vek{z})}\,.
\end{eqnarray*}
W.l.o.g.\ we have $R\neq\emptyset$. (We can assume that $R$
contains, e.g., the constraint $y \equiv_1 0$ which is satisfied by all
$\bs{y}\in\ZZ$.)\\  
We use the following fact to simplify the constraints in $R$.
\begin{fact_mit}\label{fact:residue_class}
Let $k>0$, $a_1,\twodots,a_k\in\ZZ$, $n_1,\twodots,n_k\in\NNpos$, and
let $l\deff\lcm\set{n_1,\twodots,n_k}$ \index{lcm@$\lcm$ (least common multiple)}
be the least common multiple of
$n_1,\twodots,n_k$.
\\
If there exists a $r\in\set{0,\twodots,l{-}1}$ such that 
\,$\big(r\equiv_{n_1} a_1 \, \und\, \cdots\, \und\, r\equiv_{n_k} a_k\big)$\,
then, for all $b\in\ZZ$, we have
\[
  \big(b\equiv_{n_1} a_1 \ \und\ \cdots\ \und\ b\equiv_{n_k} a_k\big)
  \qquad\mbox{iff}\qquad
  b\equiv_l r\,.
\]
If no such $r$ exists, then no $b\in\ZZ$ satisfies
\,$\big(b\equiv_{n_1} a_1 \, \und\, \cdots\, \und\, b\equiv_{n_k} a_k\big)$.
\end{fact_mit}%
\begin{proof_mit}
Let us first consider the case where there exists a $r\in\set{1,\twodots,l{-}1}$ such that 
\,$\big(r\equiv_{n_1} a_1 \, \und\, \cdots\, \und\, r\equiv_{n_k} a_k\big)$.
It is obvious that
\,$\big(b\equiv_{n_1} a_1 \, \und\, \cdots\, \und\, b\equiv_{n_k} a_k\big)$\,
if and only if
\,$\big(b\equiv_{n_1} r \, \und\, \cdots\, \und\, b\equiv_{n_k} r\big)$\, 
if and only if
\,$\big(n_1\mid b{-}r \,\und\, \cdots\,\und\, n_k \mid b{-}r\big)$\,
if and only if
\,$\lcm\set{n_1,\twodots,n_k} \mid b{-}r$\,
if and only if
\,$b\equiv_l r$.
\par
For the second claim of the above fact we prove the
contraposition. I.e.\ we assume that there is some $b\in\ZZ$
with \,$\big(b\equiv_{n_1} a_1 \, \und\, \cdots\, \und\, b\equiv_{n_k}
a_k\big)$, and we show that an appropriate $r$ does exist. In fact,
let $r\in\set{0,\twodots,l{-}1}$ be the residue class of $b$ modulo
$l$, i.e.\ $r\equiv_l b$. Since $l=\lcm\set{n_1,\twodots,n_k}$, we know that
$n_j\mid l$, and hence \,$r\equiv_{n_j} b \equiv_{n_j} a_j$, for every
$j\in\set{1,\twodots,k}$.
\end{proof_mit}
\\
\parno
The above fact tells us that we can replace the constraints in $R$ by
the single constraint $y\equiv_l \textit{res}$, where 
$l \deff \lcm\setc{n_j}{j \mbox{ such that } \alpha_j\in R}$, and
where the interpretation of the variable $\textit{res}$ is determined
by the $\FO(\Presb)$-formula
\begin{eqnarray*}
  \psi_{\textit{res}}(\textit{res},\vek{z})
& \,\deff\,
& \big(\,0\leq \textit{res} < l{\cdot} 1\,\big)
  \ \und\, 
  \Und_{j:\alpha_j \in R} \textit{res} \equiv_{n_j} t_j(\vek{z})\,.
\end{eqnarray*}
As already done in Example~\ref{example:Presb_Konjunktionen}, we 
consider a variable $\textit{first}$ which is 
interpreted by the smallest integer $> \textit{low}$ that belongs to
the correct residue class, i.e.\ that satisfies 
\,$\textit{first}\equiv_l \textit{res}$. This interpretation of the
variable $\textit{first}$ can be enforced by the $\FO(\Presb)$-formula
\begin{eqnarray*}
  \psi_{\textit{first}}(\textit{first},\textit{low},\textit{res})
& \,\deff\,
& \big(\textit{first} > \textit{low}\big)\ \und\ 
  \big(\textit{first}\equiv_l \textit{res}\big)\ \und
\\
&
& \big( \forall v\ (v>\textit{low} \,\und\, v\equiv_l \textit{res})\impl
  v\geq \textit{first}\big)
  \,.
\end{eqnarray*}
The constraints in $L\cup U\cup R$ can be visualized as shown in
Figure~\ref{figure:example:Presb_Konjunktionen}, from which we can
directly see that there are exactly
\ \( 
  \max \left\{\,0\; ,\; 
  \aufgerundet{\frac{\textit{up}-\textit{first}}{l}}
  \right\}
\) \ 
many different $\bs{y}\in\ZZ$ which satisfy all the constraints in
$L\cup U\cup R$, provided that $\textit{res}$ exists and that $L$ and
$U$ are nonempty. 
If $\textit{res}$ exists and $L$ or $U$ are empty, then there are
infinitely many $\bs{y}\in\ZZ$ which satisfy all the constraints in $L\cup
U\cup R$. If $\textit{res}$ does not exist, then no $\bs{y}\in\ZZ$ 
satisfies these constraints.
\par
Remember that the given formula \,$\varphi(x,\vek{z})\, \deff\,
\exists^{=x}y\,\Und_{j=1}^{n}\alpha_j(y,\vek{z})$\, may have, apart
from the constraints in $L\cup U\cup R$, also constraints from $I$
which are independent of $y$. (However, we assume that there are no
equations, i.e.,\ that $E=\emptyset$.)
Altogether we obtain that $\varphi(x,\vek{z})$ is equivalent to the
$\FO(\Presb)$-formula \,$\psi(x,\vek{z})\deff$
{\small
\[
\begin{array}{l}
  \displaystyle
  \Big(\,\big(\nicht\!\!\!\Und_{j:\alpha_j\in I}\!\!\! \beta_j(\vek{z})\,\big)\ \impl\ 
  x{=}0\,\Big)\ \ \und\ \ 
  \Big(\, \big(\!\!\!\Und_{j:\alpha_j\in I}\!\! \beta_j(\vek{z})\big) \ \impl\ 
\\[2ex]
  \displaystyle
  \qquad
  \big(\,  
    (\,\nicht\exists
    \textit{res}\;\psi_{\textit{res}}(\textit{res},\vek{z})\,)
    \, \impl\, 
    x{=}0
  \,\big)\ \, \und\ \,  
  \big(\,
    (\,\exists
    \textit{res}\;\psi_{\textit{res}}(\textit{res},\vek{z})\,)
    \, \impl\, \chi(x,\vek{z},\textit{res})
  \,\big)\,\Big)\,,
\end{array}
\]
}%
where $\chi$ is defined as follows:
If $L$ or $U$ are empty, then
\,$\chi(x,\vek{z},\textit{res})\,\deff\,
\nicht x{=}x$.\,
If $L$ and $U$ are nonempty, then
\,$\chi(x,\vek{z},\textit{res})\,\deff\,$
\[
\begin{array}{l}
   \exists\; \textit{low}\ \exists\; \textit{up}\ \exists\;
   \textit{first}\ 
 \\[1ex] 
   \qquad
   \psi_{\textit{low}}(\textit{low},\vek{z})\ \ \und\ \ 
   \psi_{\textit{up}}(\textit{up},\vek{z})\ \ \und\ \ 
   \psi_{\textit{first}}(\textit{first},\textit{low},\textit{res})\ \ \und
 \\[1ex] 
   \qquad
   \big(\textit{up}\leq\textit{first} \,\impl\, x{=}0\big)\ \ \und\ \  
   \big(\textit{up}>\textit{first} \,\impl\, x= \aufgerundet{\frac{\textit{up}-\textit{first}}{l}}\,\big)
   \,.
\end{array}
\]
This completes the proof of Lemma~\ref{lemma:Presb_Konjunktionen}.
\end{proofc_mit}%
\\
\parno
From Lemma~\ref{lemma:Presb_Konjunktionen} we know how to eliminate
the counting quantifier from a formula of the form
\,$\exists^{=x}y\;\theta(y,\vek{z})$, where $\theta$ is a
conjunction of $\Presb$-atoms.
The following lemma lifts 
the elimination of the counting
quantifier to be valid also for formulas where $\theta$ is a
\emph{disjunction} of conjunctions of atoms.
\begin{lemma_mit}\label{lemma:Presb_DNF}
Every $\FOunC(\Presb)$-formula of the form
\ \(
   \exists^{=x}y\; \Oder_{i=1}^{m} \Und_{j=1}^{n_i} \alpha_{i,j}(y,\vek{z})\,,
\) \ 
where the $\alpha_{i,j}$ are atoms built from the symbols in
$\set{=}\cup\Presb\cup\set{y,\vek{z}}$, 
is equivalent over $\struc{\ZZ,\Presb}$ to a $\FO(\Presb)$-formula.
\end{lemma_mit}%
\begin{proof_mit}
The proof makes use of Lemma~\ref{lemma:Presb_Konjunktionen} and of
the well-known \emph{principle of inclusion and exclusion} ($P.I.E.$\ for
short; cf., e.g., the textbook \cite{Cameron}):
\begin{fact_mit}[$\bs{P.I.E.}$]\label{fact:PIE}\index{principle of
    inclusion and exclusion}\index{PIE@P.I.E.}
Let $m>0$ and let $C_1,\twodots,C_m$ be sets.
The following is true:
\\
\centerline{$\displaystyle
   | \bigcup_{i=1}^{m} C_i |
   \quad =
  \sum_{\emptyset\neq I\subseteq\set{1,\twodots,m}} (-1)^{|I|-1}\cdot
  |\bigcap_{i\in I} C_i|
$}\\ \mbox{ }
\end{fact_mit}%
We now concentrate on the proof of Lemma~\ref{lemma:Presb_DNF}.\\
Let \,$\exists^{=x}y\; \Oder_{i=1}^{m} C_i(y,\vek{z})$ be the given
formula, where $C_i(y,\vek{z}) \deff \Und_{j=1}^{n_i}
\alpha_{i,j}(y,\vek{z})$.
Let $\boldsymbol{x,\vek{z}}$ be interpretations in $\ZZ$
of the variables $x,\vek{z}$.
We write $\boldsymbol{C_i({\cdot},\vek{z})}$ to denote the set of all
$\boldsymbol{y}\in\ZZ$ for which the conjunction
$C_i({y,\vek{z}})$ is satisfied when interpreting $y,\vek{z}$ by
$\boldsymbol{y,\vek{z}}$.
Obviously, we have
{\small\[
\begin{array}{c}
 \displaystyle
 \struc{\ZZ,\Presb,\boldsymbol{x,\vek{z}}}
 \quad\models\quad
 \exists^{=x}y\; \Oder_{i=1}^{m} C_i(y,\vek{z})
\\[1ex]
 \displaystyle
 \mbox{ iff }
\\[1ex]
 \displaystyle
 \boldsymbol{x} 
 \quad = \quad
 |\bigcup_{i=1}^m \bs{C_i({\cdot},\vek{z})}|
 \quad \stackrel{P.I.E.}{=} \quad
 {\sum_{\emptyset\neq I\subseteq \set{1,\twodots,m}}} (-1)^{|I|-1}\cdot
  |\bigcap_{i\in I} \bs{C_i({\cdot},\vek{z})}|\,.
\end{array}
\]
}For every set $I$ we
introduce a new variable $x_I$ with the intended meaning that
$x_I = |\bigcap_{i\in I} \bs{C_i({\cdot},\vek{z})}|$, which can be
enforced by the formula 
\,$\exists^{=x_I}y\; {\Und_{i\in I}} C_i(y,\vek{z})$.
This leads to the fact that
{\small
\[
\begin{array}{c}
 \displaystyle
 \boldsymbol{x} 
 \quad = \quad
 {\sum_{I}} (-1)^{|I|-1}\cdot
  |\bigcap_{i\in I} \bs{C_i({\cdot},\vek{z})}|
\\[1ex]
 \displaystyle
 \mbox{ iff }
\\[1ex]
 \displaystyle
 \struc{\ZZ,\Presb,\boldsymbol{x,\vek{z}}}
 \ \ \models\ \ 
 \big(\exists x_I\big)_{I}\;
 \big(x = {\displaystyle\sum_{I}} (-1)^{|I|-1}\cdot x_I\big)\ \und\ \  
 \Und_I \big(\,\exists^{=x_I}y\; {\displaystyle\Und_{i\in I}} C_i(y,\vek{z})\big)\,.
\end{array}
\] 
}Since \,$\Und_{i\in I} C_i(y,\vek{z}) \,=\,
\Und_{i\in I}\Und_{j=1}^{n_i} \alpha_{i,j}(y,\vek{z})$\,, we can apply
Lemma~\ref{lemma:Presb_Konjunktionen} to replace each subformula
$\exists^{=x_I}y\; {\Und_{i\in I}} C_i(y,\vek{z})$ by a
$\FO(\Presb)$-formula $\varphi_I(x_I,\vek{z})$.
\\
Altogether, we obtain a $\FO(\Presb)$-formula $\psi(x,\vek{z})$ which
is equivalent over $\struc{\ZZ,\Presb}$ to $\exists^{=x}y\;
\Oder_{i=1}^{m} C_i(y,\vek{z})$. 
This completes the proof of Lemma~\ref{lemma:Presb_DNF}.
\end{proof_mit}%
\\
\parno
The result of Theorem~\ref{maintheorem:Presb_Unary_Counting}, stating that
Presburger's quantifier elimination can be extended to unary counting
quantifiers, now is an easy consequence of
Lemma~\ref{lemma:Presb_DNF} and Theorem~\ref{theorem:Presburger}:
\medskip
\\
\begin{proofc_mit}{of Theorem
    \ref{maintheorem:Presb_Unary_Counting}}\mbox{ }\\
According to Presburger's quantifier elimination
(Theorem~\ref{theorem:Presburger}) it suffices to show that every
$\FOunC(\Presb)$-formula $\varphi$ is equivalent over
$\struc{\ZZ,\Presb}$ to a $\FO(\Presb)$-formula $\psi$.
We proceed by induction on the construction of $\varphi$.
The only nontrivial case is when $\varphi$ is of the form
\,$\exists^{=x}y\;\chi(y,\vek{z})$.
By the induction hypothesis $\chi$ is equivalent to a formula of the
form
\,$ \Oder_{i=1}^{m}\Und_{j=1}^{n_i}\alpha_{i,j}(y,\vek{z})$, where the
$\alpha_{i,j}$ are atoms built from the symbols in
$\set{=}\cup\Presb\cup\set{y,\vek{z}}$. Thus, from
Lemma~\ref{lemma:Presb_DNF} we obtain the desired $\FO(\Presb)$-formula
$\psi(x,\vek{z})$ which is equivalent to $\varphi(x,\vek{z})$ over
$\struc{\ZZ,\Presb}$. 
\end{proofc_mit}%
%
%
%
%
\subsection{From $\struc{\ZZ,\Presb}$ to $\struc{\NN,\Presb}$}
It is straightforward to transfer Theorem~\ref{maintheorem:Presb_Unary_Counting} from $\ZZ$
to $\NN$ to obtain
\begin{center}\bf
$\bs{\FOunC(<,+)\ =\ \FO(<,+)}$ \ on $\NN$. 
\end{center}
Precisely, this means:
\begin{corollary_mit}\label{corollary:Presb_Unary_Counting_NN}
Every $\FOunC(\Presb)$-formula $\varphi(\vek{z})$ is equivalent over
$\struc{\NN,\Presb}$ to a formula of the form
\ \(
   \Oder_{i=1}^{m}\Und_{j=1}^{n_i} \alpha_{i,j}(\vek{z})\,,
\) \ 
where the $\alpha_{i,j}$ are atoms built from the symbols in
$\set{=}\cup\Presb\cup\set{\vek{z}}$. 
\end{corollary_mit}%
\begin{proof_mit}
We make use of Theorem~\ref{maintheorem:Presb_Unary_Counting} and of the following 
relativization of quantifiers which gives us, for every $\FOunC(\Presb)$-formula $\varphi(\vek{z})$,
a $\FOunC(\Presb)$-formula $\tilde{\varphi}(\vek{z})$ such that the following is valid for
every interpretation $\bs{\vek{z}}\in\NN$ of the variables $\vek{z}$
\[
  (*):\qquad 
  \struc{\NN,\Presb,\bs{\vek{z}}} \models \varphi({\vek{z}})
  \quad\mbox{ iff }\quad
  \struc{\ZZ,\Presb,\bs{\vek{z}}} \models \tilde{\varphi}({\vek{z}}).
\]
The formula $\tilde{\varphi}$ is defined inductively via
\begin{displaymath}
\begin{array}{ll}
 \bullet &
 \mbox{$\tilde{\varphi} \deff\, \varphi$ \quad if\quad $\varphi$ is atomic,}
 \medskip\\
 \bullet &
 \mbox{$\tilde{\varphi} \deff\, \nicht \tilde{\psi}$ \quad if\quad $\varphi = \nicht \psi$,}
 \medskip\\
 \bullet &
 \mbox{$\tilde{\varphi} \deff\, \tilde{\psi} \oder \tilde{\chi}$ \quad if\quad
   $\varphi = \psi \oder \chi$,}
 \medskip\\
 \bullet &
 \mbox{$\tilde{\varphi} \deff\, \exists y\; \big(y\geq 0\;\und\;\tilde{\psi} \big)$ \quad if\quad
   $\varphi = \exists y\; \psi$, \quad and}
 \medskip\\
 \bullet &
 \mbox{$\tilde{\varphi} \deff\, \exists^{=x} y\; \big(y\geq 0\;\und\;\tilde{\psi} \big)$ 
    \quad if\quad $\varphi = \exists^{=x} y\; \psi$.}
\end{array}
\end{displaymath} 
It is straightforward to see that $(*)$ is indeed true.
\\
According to Theorem~\ref{maintheorem:Presb_Unary_Counting}, the formula 
$\tilde{\varphi}(\vek{z})$ is equivalent over $\struc{\ZZ,\Presb}$ to a formula of the 
form \ $\Oder_{i=1}^{m}\Und_{j=1}^{n_i}\alpha_{i,j}(\vek{z})$\,, \,where the $\alpha_{i,j}$ are
{$\Presb$-atoms}.
It is obvious that, whenever the variables $\vek{z}$ are interpreted by non-negative integers
$\bs{\vek{z}}\in\NN$, it makes no difference whether the atom $\alpha_{i,j}(\vek{z})$ is
evaluated in the structure $\struc{\ZZ,\Presb}$ or in the structure $\struc{\NN,\Presb}$.
We thus obtain for every interpretation $\bs{\vek{z}}\in\NN$ of the variables $\vek{z}$ that
\[
 \begin{array}{lrcl}
   & \struc{\NN,\Presb,\bs{\vek{z}}} & \models
   & \varphi(\vek{z})
   \medskip\\
     \mbox{iff}\quad
   & \struc{\ZZ,\Presb,\bs{\vek{z}}} & \models
   & \tilde{\varphi}(\vek{z})
   \medskip\\
     \mbox{iff}
   & \struc{\ZZ,\Presb,\bs{\vek{z}}} & \models
   & \Oder_{i=1}^{m}\Und_{j=1}^{n_i}\alpha_{i,j}(\vek{z})
   \medskip\\
     \mbox{iff}
   & \struc{\NN,\Presb,\bs{\vek{z}}} & \models
   & \Oder_{i=1}^{m}\Und_{j=1}^{n_i}\alpha_{i,j}(\vek{z})\,.
 \end{array}
\] 
I.e., the $\FOunC(\Presb)$-formula $\varphi(\vek{z})$ is equivalent over
$\struc{\NN,\Presb}$ to the quantifier free formula 
\,$\Oder_{i=1}^{m}\Und_{j=1}^{n_i} \alpha_{i,j}(\vek{z})$.
\end{proof_mit}
\subsection{From $\struc{\ZZ,\Presb}$ to Initial Segments of $\NN$}
Let us now investigate the finite versions of Presburger
arithmetic, where the universe is some initial segment of $\NN$. I.e.,
for every $N\in\NNpos$ we want to consider the  
substructure of $\struc{\NN,\Presb}$ with universe $\dom{N}\deff\set{0,\twodots,N}$.\\
There is some technical difficulty since $\dom{N}$ is not closed under the
addition function $f_+$.
We therefore move over to the version $\Presb'$ of 
Presburger arithmetic without function symbols. I.e.,
\index{Presb@$\Presb'$} 
\ \(
\Presb' \deff \set{\,0,\,1,\,R_{+},\,<,\,({\equiv_n})_{n\in\NNpos}\,},
\) \
where $R_{+}$ denotes the ternary addition relation (which, in this paper, is usually 
simply denoted $+$).
\\
Now, the initial segment
$\struc{\dom{N},{\Presb'}}$ of Presburger
arithmetic is defined in the canonical way, i.e.,
$R_+$ is interpreted by the set of all triples $(a,b,c)\in\dom{N}^3$ for which $a+b=c$. 
The aim of this section is to show that
\begin{center}\bf
$\bs{\FOunC(<,+) = \FO(<,+)}$ \ on initial segments of $\NN$.
\end{center}
Precisely, this means:
\begin{corollary_mit}\label{corollary:Presb_Unary_Counting_finite} 
Every $\FOunC(\Presb')$-formula $\varphi(\vek{z})$ is equivalent over the class \linebreak[4]
$\setc{\struc{\dom{N},{\Presb'}}}{N\in\NNpos}$\,
to a $\FO(\Presb')$-formula $\psi(\vek{z})$.
\end{corollary_mit}%
Here, the formula $\psi$ cannot be taken quantifier free in general,
because the addition is not present as a
\emph{function} (allowing to express summations of more that just
two variables), but only as a \emph{relation}.
\medskip\\
\begin{proofc_mit}{of Corollary~\ref{corollary:Presb_Unary_Counting_finite}}\mbox{}\\
The proof is similar to the proof of
Corollary~\ref{corollary:Presb_Unary_Counting_NN}. However, the fact
that the universe is finite and that the addition is only present as a
relation, causes some technical problems.
We make use of Theorem~\ref{maintheorem:Presb_Unary_Counting} and
an  appropriate relativization of quantifiers: We introduce a new variable $\Max$ with the
intended meaning that $\Max$ denotes the maximum element $N$ in the
underlying finite universe; and we transform a given
$\FOunC(\Presb')$-formula $\varphi(\vek{z})$ into an appropriate
$\FOunC(\Presb)$-formula $\tilde{\varphi}(\vek{z},\Max)$ such that the
following is valid for
every $N\in\NNpos$ and for every interpretation $\bs{\vek{z}}\in\dom{N}$ of the variables $\vek{z}$
\[
  (*):\qquad 
  \struc{\dom{N},{\Presb'},\bs{\vek{z}}} \models \varphi({\vek{z}})
  \quad\mbox{ iff }\quad
  \struc{\ZZ,\Presb,\bs{\vek{z}},N} \models \tilde{\varphi}({\vek{z}},\Max).
\]
The formula $\tilde{\varphi}$ is defined inductively via
\begin{displaymath}
\begin{array}{ll}
 \bullet &
 \mbox{$\tilde{\varphi} \deff\, f_+(x,y) = z$ \quad if\quad $\varphi = R_+(x,y,z)$,}
 \medskip\\
 \bullet &
 \mbox{$\tilde{\varphi} \deff\, \varphi$ \quad if\quad $\varphi$ is
 an atom not involving the addition relation $R_+$,}
 \medskip\\
 \bullet &
 \mbox{$\tilde{\varphi} \deff\, \nicht \tilde{\psi}$ \quad if\quad $\varphi = \nicht \psi$,}
 \medskip\\
 \bullet &
 \mbox{$\tilde{\varphi} \deff\, \tilde{\psi} \oder \tilde{\chi}$ \quad if\quad
   $\varphi = \psi \oder \chi$,}
 \medskip\\
 \bullet &
 \mbox{$\tilde{\varphi} \deff\, \exists y\; \big(0\leq y\leq \Max\;\und\;\tilde{\psi} \big)$ \quad if\quad
   $\varphi = \exists y\; \psi$, \quad and}
 \medskip\\
 \bullet &
 \mbox{$\tilde{\varphi} \deff\, \exists^{=x} y\; \big(0\leq y \leq \Max\;\und\;\tilde{\psi} \big)$ 
    \quad if\quad $\varphi = \exists^{=x} y\; \psi$.}
\end{array}
\end{displaymath} 
It is straightforward to see that $(*)$ is indeed true.
\\
According to Theorem~\ref{maintheorem:Presb_Unary_Counting}, the formula 
$\tilde{\varphi}(\vek{z},\Max)$ is equivalent over $\struc{\ZZ,\Presb}$ to a formula of the 
form \ $\Oder_{i=1}^{m}\Und_{j=1}^{n_i}\alpha_{i,j}(\vek{z},\Max)$\,, \,where the $\alpha_{i,j}$ are
atoms built from the symbols in
$\set{=}\cup\Presb\cup\set{\vek{z},\Max}$. 
\\
Of course, it suffices to show that each such $\Presb$-atom can be transformed
into a $\FO(\Presb')$-formula $\chi_{i,j}(\vek{z},\Max)$ such that
the following is valid for every $N\in\NNpos$ and for every
interpretation $\bs{\vek{z}}\in\dom{N}$ of the variables $\vek{z}$
{\small
\[
  ({**}):\quad  
  \struc{\ZZ,\Presb,\bs{\vek{z}},N} \models  \alpha_{i,j}({\vek{z}},\Max)
  \quad\mbox{iff}\quad
  \struc{\dom{N},{\Presb'},\bs{\vek{z}},N} \models \chi_{i,j}({\vek{z}},\Max).
\]}%
To see what is the problem about defining $\chi_{i,j}$, let us have a
closer look at the $\Presb$-atom $\alpha_{i,j}$.
By definition, the atom $\alpha_{i,j}$ expresses that
\begin{eqnarray*}
  ({**}*):\qquad u_1 + \cdots + u_k & \ltimes & v_1 + \cdots + v_l
\end{eqnarray*}
where $k,l\in\NN$, \,${\ltimes}$ is an element in $\setc{=,<,\equiv_n}{n\in\NNpos}$, \,and
$u_1,\twodots,u_k,v_1,\twodots,v_l$ are (not necessarily distinct)
elements in $\set{0,1,\vek{z},\Max}$.
\\
When the variable $\Max$ is interpreted by some $N\in\NNpos$ and the
variables $\vek{z}$ are interpreted by numbers
$\bs{\vek{z}}\in\dom{N}$, then the term $u_1+\cdots +u_k$ evaluates to
a number of size at most $k\cdot N$. But there is no guarantee that
this number does not exceed $N$, i.e. that it belongs to the
underlying finite universe $\dom{N}$. We therefore have to move over to the
$2$-tuple version 
which allows us to
represent a number $x$ of size at most $N^2$ ($> k{\cdot}N$, for $N$
large enough) by two
numbers $x_1,x_0$ in $\dom{N}$ via $x= x_1{\cdot}(N{+}1) +
x_0$. From Theorem~\ref{theorem:d-arithmetic} we know that there is a
$\FO(R_+)$-formula $\varphi_{+}^2$ which expresses the addition
relation for numbers that are represented by such $2$-tuples. 
Hence, the result of the summation 
\ \(
  u_1\, + \,\cdots\, + \,u_k 
\) \ 
is the number represented by the 2-tuple $(u_1^{(k)},u_0^{(k)})$,
where $(u_1^{(k)},u_0^{(k)})$ is determined by the $\FO(R_+)$-formula
{\small
\[
   \big(\exists u_1^{(i)}\, \exists u_0^{(i)}\big)_{i\leq k}\ 
   u_1^{(1)}{=}0
   \;\und\; 
   u_0^{(1)}{=} u_1 
   \;\und\,
   \Und_{i=2}^{k} \varphi_+^2\big(u_1^{(i-1)}, u_0^{(i-1)},0,u_i,u_1^{(i)},u_0^{(i)} \big)\,.
\]
}%
In the same way one obtains a 2-tuple $(v_1^{(l)},v_0^{(l)})$ which
represents the result of the summation \,$v_1+\cdots+v_l$.
Finally, making use of the formulas $\varphi^2_<$ and
$\varphi^2_{\equiv_n}$ from Theorem~\ref{theorem:d-arithmetic}, $({**}*)$ can be replaced by 
\begin{itemize}
 \item 
   $\big(u_1^{(k)} {=} v_1^{(l)} \ \und\ u_0^{(k)} {=} v_0^{(l)}\big)$ \ \ if \ 
   $\ltimes$ is $=$
 \item 
   $\varphi^2_<\big(u_1^{(k)},u_0^{(k)},v_1^{(l)},v_0^{(l)}\big)$ \ \ if \ 
   $\ltimes$ is $<$
 \item 
   $\varphi^2_{\equiv_n}\big(u_1^{(k)},u_0^{(k)},v_1^{(l)},v_0^{(l)}\big)$ \ \ if \ 
   $\ltimes$ is a congruence relation $\equiv_n$.
\end{itemize}
Altogether we obtain a $\FO(\Presb')$-formula
$\chi_{i,j}$ that has property $(**)$. This completes the proof
of Corollary~\ref{corollary:Presb_Unary_Counting_finite}.
\end{proofc_mit}
%
%



\section{Applications}\label{section:Applications}
In this section we point out some applications of the 
previous section's result that 
$\FOunC(<,+)=\FO(<,+)$ on $\NN$ and on initial segments of $\NN$.
We obtain the failure of a particular version of the Crane Beach conjecture, and 
we show that reachability and
connectivity of finite graphs are not expressible in first-order logic with unary counting
and addition.
\subsection{Reachability}
A finite \emph{graph} \index{graph}
$G=\struc{V,E}$ consists of a finite set $V$ of
\emph{vertices} and a set $E\subseteq V^2$ of directed edges. 
A directed \emph{path} $p= (v_0,\twodots,v_k)$ of length $k$ is a \index{path}
sequence of vertices  satisfying $E(v_{i},v_{i+1})$ for all
$i\in\set{0,\twodots,k{-}1}$. 
We say that $p$ is a path from $s$ to $t$ if $s=v_0$ and $t=v_k$.
The path $p$ is called \emph{deterministic} \index{deterministic path}
if, for
each $i\in\set{0,\twodots,k{-}1}$, the
edge $(v_i,v_{i+1})$ is the \emph{unique} edge in $G$ leaving vertex
$v_i$, i.e., there is no vertex $u\neq v_{i+1}$ in $V$ such that $E(v_i,u)$.
\begin{definition_mit}[$\detReach$]\label{definition:reachability}\index{reachability}\index{deterministic reachability}\index{DetReach@$\detReach$}  
\mbox{ }\\
\emph{Deterministic Reachability}, $\detReach$, is the set of all finite
graphs $G$, together with a distinguished source vertex $s$ and a
distinguished target vertex $t$, such that there is a \emph{deterministic
  path} from $s$ to $t$ in $G$.
I.e.,
{\small
\[
 \detReach \, \deff \,
 \bigsetc{\,\struc{G,s,t}}{
   \begin{array}{ll} 
     G=\struc{V,E} \mbox{ is a finite graph},\ s,t\in V, \mbox{ and
     }
   \\
     \mbox{there is a deterministic path from $s$ to $t$ in $G$}
   \end{array}
  }\,.
\]
}%
\end{definition_mit}%
The problem $\detReach$ is complete for $\Logspace$ \index{LOGSPACE@$\Logspace$} via 
first-order reductions (cf., e.g., the textbook
\cite[Theorem~3.23]{Immerman}).
An important open question in complexity theory is to separate
$\Logspace$ from other, potentially weaker, complexity classes. Such a
separation could be achieved by showing that the problem $\detReach$
does not belong to the potentially weaker class.
One potentially weaker class for which no separation from $\Logspace$
is known by now\footnote{$\TCzero$ has not even been separated from
  $\NP$, cf., \cite{Allender}.}, is
the class $\TCzero$ \index{TC0@$\TCzero$} that
consists of all problems solvable with uniform 
threshold circuits of polynomial size and constant depth.
As already mentioned in Section~\ref{section:Presb_and_Counting:known},
it was shown in \cite{BIS} that, for ordered structures over
arbitrary signatures $\tau$, 
logtime-uniform $\TCzero$ is exactly the class of all problems definable in 
$\FOunC(+,\times,\tau)$.
To separate $\TCzero$ from $\Logspace$ it would therefore suffice to
show that \,$\detReach$ is not definable in $\FOunC(+,\times,E,s,t)$.
\\
%
Ruhl \cite{Ruhl} achieved a first step towards such a separation by
showing the (weaker) result that
%
${\detReach}$ is not definable in ${\FOunC(+,E,s,t)}$. 
Precisely, this means:
\begin{theorem_mit}[Ruhl \cite{Ruhl}]\label{theorem:Ruhl}\index{Ruhl, Matthias}
There is no $\FOunC(+,E,s,t)$-sentence $\psi$ such that, for every
$N\in\NNpos$ and all graphs $G=\struc{\dom{N},E}$ with vertex set
$\dom{N}$, and vertices $s,t\in \dom{N}$, we have
\ \(
\struc{G,s,t}\in \detReach
\ \mbox{ iff } \
\struc{\dom{N},+,E,s,t} \models \psi.
\)
\end{theorem_mit}%
The aim of this section is to point out that Ruhl's theorem can be
proved easily when making use of our 
result that {${\FOunC(+) = \FO(+)}$ on initial segments of ${\NN}$}.
Before presenting the easy proof, let us first outline Ruhl's
approach:
\par
\begin{figure}[!htbp]
\bigskip
\begin{center}
\fbox{
\scalebox{0.65}{
\begin{pspicture}(-1,-2.75)(+16,2.75)
\rput(0,0){\rnode{nullnull}{\large{$\mid$}}}%
\rput(0,-0.5){\rnode{null}{\large$0$}}%
\rput(3,0){\rnode{RR}{\large{$\mid$}}}%
\rput(3,-0.5){\rnode{R}{\large$R$}}%
\rput(15,0){\rnode{NN}{\large{$\mid$}}}%
\rput(15,-0.5){\rnode{N}{\large$N$}}%
\pnode(0,0){Links}%
\pnode(15,0){Rechts}%
\ncline[linestyle=dashed,linewidth=0.2pt]{-}{Links}{Rechts}%
\psset{fillcolor=black,linecolor=black,dotsize=5pt}%
\psdot(0,0)\psdot(1,0)\psdot(2,0)\psdot(3,0)\psdot(4,0)\psdot(5,0)%
\psdot(6,0)\psdot(7,0)\psdot(8,0)\psdot(9,0)\psdot(10,0)\psdot(11,0)%
\psdot(12,0)\psdot(13,0)\psdot(14,0)\psdot(15,0)%
\psset{arrowsize=6pt}%
\parabola{->}(0,0)(1.5,2)%
\parabola{->}(3,0)(4.5,2)%
\parabola{->}(6,0)(7.5,2)%
\parabola{->}(9,0)(10.5,2)%
\parabola{->}(12,0)(13.5,2)%
\parabola{->}(1,0)(2.5,-2)%
\parabola{->}(4,0)(5.5,-2)%
\parabola{->}(7,0)(8.5,-2)%
\parabola{->}(10,0)(11.5,-2)%
\parabola{->}(2,0)(3.5,-2)%
\parabola{->}(5,0)(6.5,-2)%
\parabola{->}(8,0)(9.5,-2)%
\parabola{->}(11,0)(12.5,-2)%
\end{pspicture}
}
}
\caption{\small Visualization of the graph $G_{R,N}$ (for $R=3$ and $N=15$). There is an edge
  from a vertex $u$ to a vertex $v$ iff $u+R=v$.}\label{figure:Ruhl}
\end{center}
\end{figure}
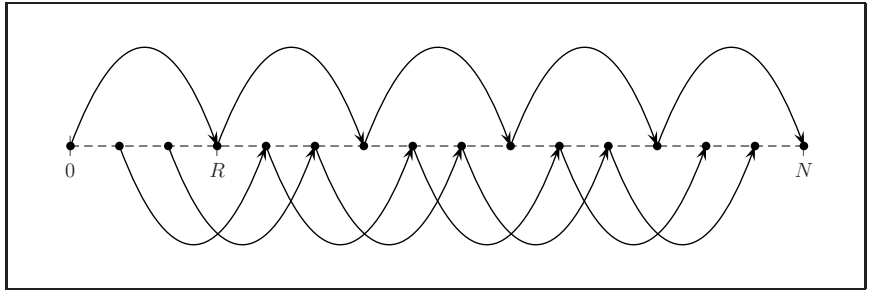

Ruhl's proof method is the Ehrenfeucht-Fra\"\i{}ss\'{e} game \index{Ehrenfeucht-Fra\"\i{}ss\'{e} game}
for
$\FOunC(+,E)$. He considers, for each $N\in\NNpos$ and $R\in\dom{N}$ the
graph $G_{R,N} = \struc{\dom{N},E_{R,N}}$ where the edge relation
$E_{R,N}$ is defined via \,``$E_{R,N}(u,v)$ iff $u+R=v$'',\, for all vertices
$u,v\in\dom{N}$.
An illustration of the graph $G_{R,N}$ is given in Figure~\ref{figure:Ruhl}.
\\
Note that the graph $G_{R,N}$ is constructed in such a way that $R\mid N$ if and
only if there is a deterministic path from $0$ to $N$, i.e.\ 
$\struc{G_{R,N},0,N}\in\detReach$.
\\
For every fixed number $M$ of rounds in the Ehrenfeucht-Fra\"\i{}ss\'{e}
game, Ruhl constructs an $N_M\in\NNpos$ and an
$R_M\in\dom{N_M}$ such that $R_M \mid N_M$ and $2R_M \nmid N_M$. 
Afterwards he explicitly exposes a clever and very intricate winning
strategy for the duplicator in the $M$-round Ehrenfeucht-Fra\"\i{}ss\'{e}
game for $\FOunC(+,E,s,t)$ on the structures $\struc{G_{R_M,N_M},0,N_M}$ and
$\struc{G_{2R_M,N_M},0,N_M}$.
He thus obtains, for every $M$, that the structures $\struc{G_{R_M,N_M},0,N_M}
\in\detReach$ and $\struc{G_{2R_M,N_M},0,N_M}\not\in\detReach$ cannot be
distinguished by {$\FOunC(+,E,s,t)$}-formulas of quantifier depth $M$.
Altogether, this
shows that $\detReach$ is not definable in $\FOunC(+,E,s,t)$.
\par
As usual in Ehrenfeucht-Fra\"\i{}ss\'{e} arguments, precise
bookkeeping is necessary for the proof.
This bookkeeping can be avoided when using
Presburger's quantifier elimination and its extension
to unary quantifiers:
\\
\parno
\begin{proofc_mit}{of Theorem~\ref{theorem:Ruhl}}
By contradiction. \\
Suppose that $\psi$ is a
$\FOunC(+,E,s,t)$-sentence defining $\detReach$. \par
The first step of the proof is to transform $\psi$ into a
$\FOunC(+)$-formula
$\varphi_1(x)$ which expresses that $R \mid N$, whenever the variable
$x$ is interpreted by a number $R$ in an underlying universe
$\dom{N}$. For this tranformation we make use of Ruhl's graphs
$G_{R,N}$ from which we know that $\struc{G_{R,N},0,N}\in\detReach$ if
and only if $R\mid N$.
The formula $\varphi_1(x)$ is obtained from $\psi$ by replacing every
atom $E(u,v)$ with the atom \linebreak[4] $u+x=v$, by replacing $s$ with $0$, and by
replacing $t$ with $\Max$. Here, $x$ is a variable not occurring in
$\psi$, and $\Max$ is a variable that is enforced to be interpreted
with the maximum element in the underlying universe.
Of course we have for all $N\in\NNpos$ and all $R\in\dom{N}$ that
\[
 \begin{array}{rcl}
   R\mid N
 & \mbox{ iff }
 & \struc{G_{R,N},0,N}\ \in\ \detReach
 \\[0.5ex]
 & \mbox{ iff }
 & \struc{\dom{N},+,E_{R,N},0,N}\ \models\ \psi
 \\[0.5ex]
 & \mbox{ iff }
 & \struc{\dom{N},+,R}\ \models\ \varphi_1(x)\,.
 \end{array}
\]
This completes the first step of the proof.
\par
From Corollary~\ref{corollary:Presb_Unary_Counting_finite} we know
that the counting quantifiers can be eliminated from $\varphi_1(x)$. I.e.,
we obtain a $\FO(+)$-formula $\varphi_2(x)$ which expresses that $R \mid
N$ whenever the variable $x$ is interpreted by a number $R$ in an underlying universe
$\dom{N}$. This gives us a $\FO(+)$-sentence $\varphi_3 \deff \forall
x\,\varphi_2(x)\impl (x{=}1 \,\oder\, x{=}\Max)$ which expresses
that $N$ is a prime number. In other words: $\varphi_3$
is a $\FO(+)$-sentence whose spectrum is the set of prime numbers.
This is a contradiction
to the Theorem of Ginsburg and Spanier (cf., 
Section~\ref{subsection:expressive_power}) and completes 
the proof of Theorem~\ref{theorem:Ruhl}.
\end{proofc_mit}%
%
%
%

%
%
\subsection{Connectivity}
\begin{definition_mit}[$\Conn$]\label{definition:connectivity}\index{connectivity}\index{Conn@$\Conn$} 
\emph{Connectivity}, $\Conn$, is the set of all finite graphs $G$
which are \emph{connected}, i.e., where there is a path from $u$ to
$v$, for all vertices $u\neq v$ in $G$.
\end{definition_mit}%
Note that Theorem~\ref{theorem:Ruhl} in particular implies that the
general reachability problem $\Reach$ (which is defined in the same
way as $\detReach$ except for replacing ``deterministic path'' with ``path'') 
is not definable in $\FOunC(+,E,s,t)$.
When considering a logic that is closed under
universal quantification, then $\Conn$ is definable as soon as
$\Reach$ is, via
\,$\psi_{\Conn} \deff \forall x\,\forall
y\,\varphi_{\Reach}(x,y)$, where $\varphi_{\Reach}$ is a formula
defining $\Reach$.
However, \emph{un}definability of $\Conn$ does not a priori follow from 
undefinability of $\Reach$. 
Nevertheless, a variation of the proof of
Theorem~\ref{theorem:Ruhl} leads to the result that also 
${\Conn}$ is not definable in ${\FOunC(+,E)}$. 
More precisely:
\begin{theorem_mit}
There is no $\FOunC(+,E)$-sentence $\psi$ such that, for every
$N\in\NNpos$ and all graphs $G=\struc{\dom{N},E}$, we have
\ \(
\struc{\dom{N},E}  \in \Conn
\ \mbox{ iff } \ 
\struc{\dom{N},+,E} \models \psi.
\)
\end{theorem_mit}%
\begin{proof_mit}
By contradiction. \\
Suppose that $\psi$ is a
$\FOunC(+,E)$-sentence defining $\Conn$. 
Again, we transform $\psi$ into a
$\FOunC(+)$-formula
$\varphi_1(x)$ which expresses that $R \mid N$, whenever the variable
$x$ is interpreted by a number $R$ in an underlying universe
$\dom{N}$. 
Instead of the graphs $G_{R,N}$ we now consider the graphs
$H_{R,N}$ illustrated and defined in Figure~\ref{figure:Conn}.
\begin{figure}[!htbp]
\bigskip
\begin{center}
\fbox{
\scalebox{0.65}{
\begin{pspicture}(-1,-2.95)(+16,2.55)
\rput(0,0){\rnode{nullnull}{\large{$\mid$}}}%
\rput(0,-0.5){\rnode{null}{\large$0$}}%
\rput(3,0){\rnode{RR}{\large{$\mid$}}}%
\rput(3,-0.5){\rnode{R}{\large$R$}}%
\rput(15,0){\rnode{NN}{\large{$\mid$}}}%
\rput(15.2,-0.5){\rnode{N}{\large$N$}}%
\pnode(0,0){Links}%
\pnode(15,0){Rechts}%
\ncline[linestyle=dashed,linewidth=0.2pt]{-}{Links}{Rechts}%
\psset{fillcolor=black,linecolor=black,dotsize=5pt}%
\psdot(0,0)\psdot(1,0)\psdot(2,0)\psdot(3,0)\psdot(4,0)\psdot(5,0)%
\psdot(6,0)\psdot(7,0)\psdot(8,0)\psdot(9,0)\psdot(10,0)\psdot(11,0)%
\psdot(12,0)\psdot(13,0)\psdot(14,0)\psdot(15,0)%
\psset{arrowsize=5pt}%
\parabola{-}(0,0)(1.5,2)%
\parabola{-}(3,0)(4.5,2)%
\parabola{-}(6,0)(7.5,2)%
\parabola{-}(9,0)(10.5,2)%
\parabola{-}(12,0)(13.5,2)%
\parabola{-}(1,0)(2.5,-2)%
\parabola{-}(4,0)(5.5,-2)%
\parabola{-}(7,0)(8.5,-2)%
\parabola{-}(10,0)(11.5,-2)%
\parabola{-}(2,0)(3.5,-2)%
\parabola{-}(5,0)(6.5,-2)%
\parabola{-}(8,0)(9.5,-2)%
\parabola{-}(11,0)(12.5,-2)%
\pscurve{-}(1,0)(1.55,-0.5)(2,0)%
\pscurve[curvature=0.6 0.6 0]{-}(1,0)(1.9,-2.1)(5,-2.4)%
\psline{-}(5,-2.4)(11,-2.4)%
\pscurve[curvature=0.6 0.6 0]{-}(11,-2.4)(14.1,-2.1)(15,0)%
%
\end{pspicture}
}
}
\caption{\small Visualization of the graph $H_{R,N}$ (for $R=3$ and $N=15$). There is an edge
  between vertices $u$ and $v$ iff \ $(u{+}R{=}v)\ \oder\ (u{=}1\,\und\,
  v{=}N)\ \oder\ (0{<}u\,\und\, u{+}1{=}v\,\und\, v{<}R)$.}\label{figure:Conn}
\end{center}
\end{figure}
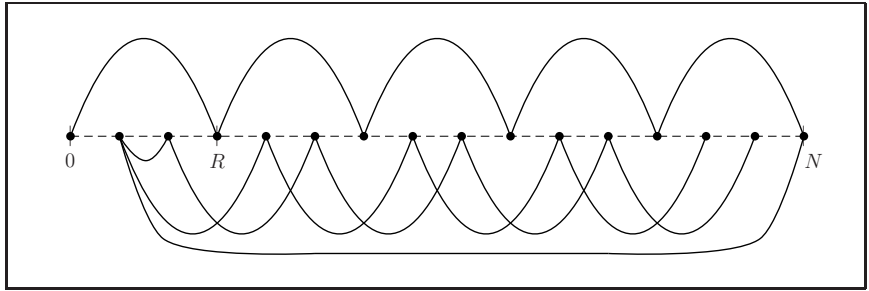
\\
It is straightforward to see that $H_{R,N}$ is connected if and only
if $R\mid N$.
%
%
\\
The formula $\varphi_1(x)$ is obtained from $\psi$ by replacing every
atom $E(u,v)$ with the formula $\chi(u,v)\oder\chi(v,u)$, where
\begin{eqnarray*}
  \chi(u,v) & \deff &
  u{+}x{=}v \ \oder\ (u{=}1 \,\und\, v{=}\Max)\ \oder\ 
  (0{<}u \,\und\, u{+}1{=}v \,\und\, v{<}x)\,.
\end{eqnarray*}
Here, $x$ is a variable not occurring in
$\psi$, and $\Max$ is a variable that is enforced to be interpreted
with the maximum element in the underlying universe.
For all $N\in\NNpos$ and all $R\in\dom{N}$ we have
\,$R\mid N$ \ iff \ $\struc{H_{R,N},+} \models \psi$ \ iff \ 
$\struc{\dom{N},+,R} \models \varphi_1(x)$.\,
The rest can be taken verbatim from the proof of Theorem~\ref{theorem:Ruhl}.
\end{proof_mit}
\subsection{A Specific Case of the Crane Beach Conjecture}
The Crane Beach conjecture deals with logical definablility of neutral letter languages. 
A language $L$ (i.e., a set of finite strings) over an alphabet $A$ is said to have
a \emph{neutral letter} $e\in A$ if inserting or deleting $e$'s from any string over
$A$ does not change the string's membership or non-membership in $L$.
Given a logic $F$ and a class $\frak{A}$ of arithmetic predicates, the Crane Beach
conjecture is said to be true for $F(<,\frak{A})$ iff for every finite alphabet $A$ and 
every neutral letter language $L$ over $A$ the following is true: If $L$ is definable in
$F(<,\frak{A})$ then $L$ is already definable in $F(<)$. 
\par
The Crane Beach conjecture is closely related to uniformity conditions in circuit complexity
theory and to collapse results in database theory.
Depending on the logic $F$ and the predicates $\frak{A}$, the Crane Beach conjecture turns
out to be true for some cases and false for others.
A detailed investigation and a state-of-the-art overview of what is known about 
the Crane Beach conjecture can be found in \cite{BILST-full}.
Using the result of Corollary~\ref{corollary:Presb_Unary_Counting_NN} that 
$\FOunC(<,+)=\FO(<,+)$ on $\NN$, one can prove the conjecture to be false for the 
following specific case:
\begin{theorem_mit}[\mbox{\cite[Theorem 6.4\,(b)]{BILST-full}}]\label{theorem:CBC}\mbox{}\\
Let $P\subseteq\NN$ be a set that is not semi-linear.\footnote{A set $P\subseteq \NN$ is 
semi-linear iff there are $p,N_0\in\NN$ such that for every $N> N_0$ we have $N\in P$ iff 
$N{+}p \in P$.}
The Crane Beach conjecture is false for $\FOunC(<,P)$.
More precisely: There is a neutral letter language $L$ over the alphabet 
$\set{a,e}$ that can be defined in $\FOunC(<,P)$, but not in $\FOunC(<)$.
\end{theorem_mit}%
\begin{proofc_mit}{(Sketch)}
Choose $L$ to be the set of all strings $w$ over $\set{a,e}$ where the number of $a$'s in $w$
belongs to $P$. Now assume, for the sake of contradiction, that $L$ is definable by a
$\FOunC(<)$-sentence $\chi$. I.e., for every string $w$ over $\set{a,e}$ we assume that
$w$ satisfies $\chi$ iff $w\in L$. It is not difficult to translate $\chi$ into a
$\FOunC(<)$-formula $\varphi(x)$ such that 
$P = \setc{N\in\NN}{\struc{\NN,<}\models \varphi(N)}$.
However, due to Corollary~\ref{corollary:Presb_Unary_Counting_NN}, $\varphi(x)$ is equivalent over $\NN$ to a
$\FO(<,+)$-formula $\psi(x)$. 
From the Theorem of Ginsburg and Spanier we therefore obtain that $P$ is semi-linear, which
is a contradiction to the choice of $P$.
\end{proofc_mit}




\section{Conclusion and Open Questions}\label{section:Conclusion}
In this paper we have gained a complete picture of the expressive power of first-order logic
with counting quantifiers and arithmetic. This picture if visualized in the following 
Figures~\ref{figure:Inclusion_Diagram_pure} and 
\ref{figure:Inclusion_Diagram_finite_structures}.
Concerning these diagrams, the only question that remains open is whether 
$\FOunC(+,\times) = \FO(+,\times)$ on initial segments of $\NN$. As pointed out in 
Section~\ref{section:Preliminaries:Arithmetic:Pure}, inequality would imply 
$\LinH\neq\ETIME$, whereas no such complexity theoretic consequence is known for the case
of equality.
\par
A main theorem of this paper is that Presburger arithmetic is closed under 
unary counting quantifiers. As applications of this we obtained an easy proof of Ruhl's
result that deterministic reachabiliy in finite graphs 
is not definable in $\FOunC(+)$, that connectivity of finite graphs is not definable in
$\FOunC(+)$, and that the Crane Beach conjecture is false for $\FOunC(<,P)$ whenever $P$ is
a set of natural numbers that is not semi-linear.
%
%
\begin{figure}[!htbp]
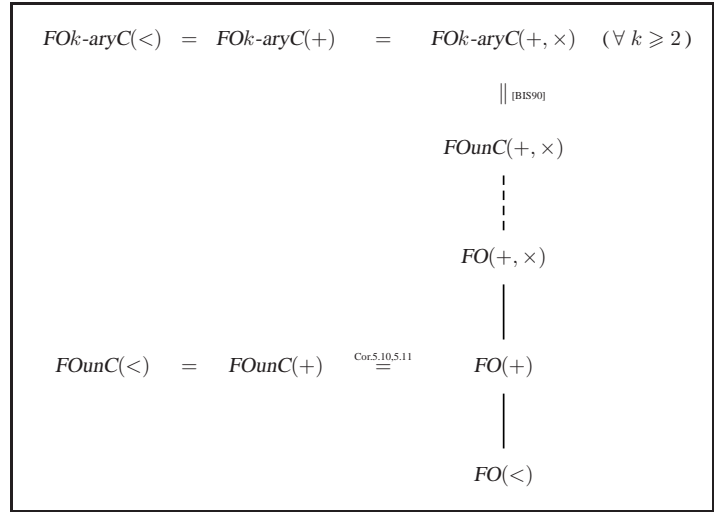

\bigskip
\begin{center}
\fbox{
\scalebox{0.80}{
\psset{framesep=7pt,labelsep=3pt,linecolor=black,linestyle=none}%
\begin{psmatrix}[rowsep=25pt,colsep=1pt]
   \psframebox{$\FOkaryC(<)$} & $=$%
 & \psframebox{$\FOkaryC(+)$} & $=$%
 & \psframebox{$\FOkaryC(+,\times)$} & & \ (\,$\forall\;k\geq 2$\,) 
 & \mbox{ \ }%
 \\%
   & %
 & & %
 & \psframebox{$\FOunC(+,\times)$}%
 \\
   & %
 & & %
 & \psframebox{$\FO(+,\times)$}
 \\
   \psframebox{$\FOunC(<)$} & $=$%
 & \psframebox{$\FOunC(+)$} & $\stackrel{\mbox{
\tiny Cor.\ref{corollary:Presb_Unary_Counting_NN},\ref{corollary:Presb_Unary_Counting_finite}}}{=}$%
 & \psframebox{$\FO(+)$}
 \\
   & %
 & & %
 & \psframebox{$\FO(<)$}%
\psset{linestyle=solid}%
\ncline{5,5}{4,5}%
\ncline{4,5}{3,5}%
\psset{linestyle=dashed}%
\ncline{3,5}{2,5}%
\psset{linestyle=none}
\ncline{2,5}{1,5}>{\tiny
\cite{BIS}}\ncput{$\parallel$}%
\end{psmatrix}%
}
}
\caption{\small Expressive power for \emph{pure arithmetic}. Lines
  indicate proper inclusions. For $\NN$ the dashed line can be
  replaced by equality (cf., Proposition~\ref{proposition:counting_NN}). 
  For initial segments of $\NN$ this remains open; however, as pointed out in
  Section~\ref{section:Preliminaries:Arithmetic:Pure}, inequality would imply 
  $\LinH\neq\ETIME$.}\label{figure:Inclusion_Diagram_pure}
\end{center}
\end{figure}%
%
%
%
\begin{figure}[!htbp]
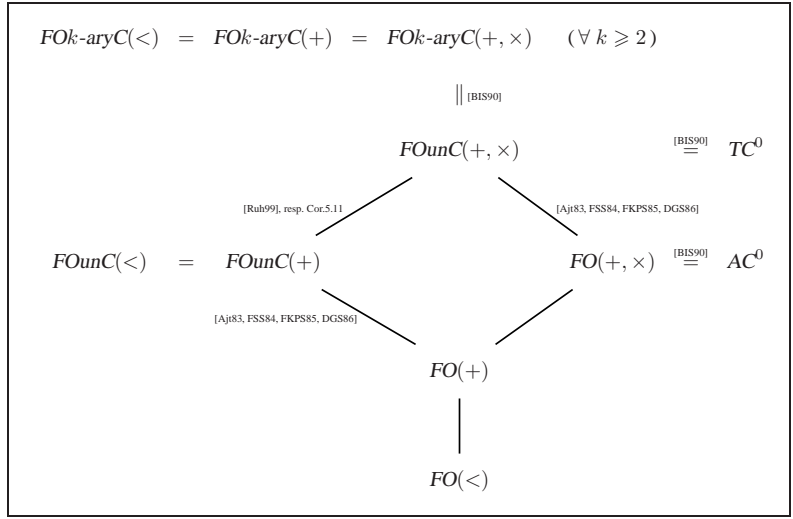

\bigskip
\begin{center}
\fbox{
\scalebox{0.80}{
\psset{framesep=7pt,labelsep=3pt,linecolor=black,linestyle=none}%
\begin{psmatrix}[rowsep=25pt,colsep=1pt]
   \psframebox{$\FOkaryC(<)$} & $=$%
 & \psframebox{$\FOkaryC(+)$} & $=$%
 & \psframebox{$\FOkaryC(+,\times)$} & & (\,$\forall\;k\geq
 2$\,) %
 \\%
   & %
 & & %
 & \psframebox{$\FOunC(+,\times)$} & & &
 $\stackrel{\mbox{
\tiny\cite{BIS}}}{=}$%
 & \psframebox{$\TCzero$}
 \\
   \psframebox{$\FOunC(<)$} & $=$%
 & \psframebox{$\FOunC(+)$} & & & %
 & \psframebox{$\FO(+,\times)$} &  $\stackrel{\mbox{
\tiny\cite{BIS}}}{=}$%
 & \psframebox{$\ACzero$}%
 \\
   & %
 & & %
 & \psframebox{$\FO(+)$}%
 \\
   & %
 & & %
 & \psframebox{$\FO(<)$}%
\psset{linestyle=solid}%
\ncline{5,5}{4,5}%
\ncline{4,5}{3,3}<{
\tiny\cite{Ajtai,FSS,FKPS,DeGuSh}}%
\ncline{4,5}{3,7}%
\ncline{3,3}{2,5}<{
\tiny\cite{Ruhl}, resp.\ Cor.\ref{corollary:Presb_Unary_Counting_finite}}%
\ncline{3,7}{2,5}>{
\tiny\cite{Ajtai,FSS,FKPS,DeGuSh}}%
\psset{linestyle=none}
\ncline{2,5}{1,5}>{
\tiny\cite{BIS}}\ncput{$\parallel$}%
\end{psmatrix}%
}
}
\caption{\small Expressive power on \emph{finite structures}. Lines
  indicate proper inclusions.}\label{figure:Inclusion_Diagram_finite_structures}
\end{center}
\end{figure}%
%
\\
\parno
With regard to the questions here, we want to mention the following interesting problems:
\begin{enumerate}[$\bullet$]
\item
  Is there any serious complexity theoretic consequence for the case that
  $\FOunC(+,\times)$ $ = \FO(+,\times)$ on initial segments of $\NN$? 
\item
  The author thinks it would be fruitful to translate
  the tools developed for bounded arithmetic 
  (cf., \cite{EsbelinMore}) into the language
  used for descriptive complexity and circuit complexity, i.e., into
  results about $\FO(+,\times)$ on initial segments of $\NN$ and on finite
  structures. 
\item
  It is an important task to search for inexpressibility results for $\FOunC(+,\times)$ 
  on finite structures. Since $\FOunC(+,\times)$ corresponds to the
  complexity class $\TCzero$, an inexpressibility result would give us a separation of 
  complexity classes. This indicates that this task will be very hard to attack.
\item
  More tractable seems the investigation of the fine
  structure of first-order logic with majority quantifiers. As 
  mentioned in Section~\ref{section:Presb_and_Counting:known},  
  it was shown in \cite{BIS} that 
  \begin{quote}\bf
    $\bs{\FObinM(<) \,=\, \FOunM(+,\times) \,=\, \FOunC(+,\times) \,=\,
    \TCzero}$\\
    on finite structures.
  \end{quote}
  In \cite{LMSV} it was shown that \,$\FOunM(<)\subsetneq \FObinM(<)$.
  It is not difficult to see that $\FO(<)\subsetneq \FOunM(<)$: The
  $\FOunM(<)$-formula
  \,\(\exists x\big( My\,(y\leq x)\,\und\, My\,(y\geq x)\big)\)\,
  expresses that the cardinality of the underlying universe is odd,
  whereas this is not expressible in $\FO(<)$ (cf., e.g., the textbook
  \cite[Example~2.3.6]{EbbinghausFlum}). \\
  It remains open whether on finite structures we have \index{FOunM<@$\FOunM(<)$}
  {\bf
  \[ \bs{
    \FOunM(<) \quad 
    \stackrel{\textrm{?}}{=} \quad 
    \FOunM(<,+) \quad 
    \stackrel{\textrm{?}}{=} \quad
    \FOunC(<,+).}
  \]}
\end{enumerate}%



%
%
%



\section*{Appendix}
\begin{proofc_mit}{of Theorem~\ref{theorem:d-arithmetic}}\mbox{}\\
For a tuple ${\vek{x}}\deff ({x_{d-1},\twodots,x_0}) \in (\dom{N})^d$ we write
$\Nr_N({\vek{x}})$ to denote the number 
$\sum_{i=0}^{d-1}{x_i}(N{+}1)^i$.
\bigskip\\ 
{\bf\em (a):}\quad Obviously, the formulas $\varphi_{<}^d$ can be inductively defined via
{\small
\[
  \begin{array}{rcl}
    \varphi_{<}^1(x_0,y_0) 
  & \deff 
  & x_0{<}y_0\,, \quad \mbox{ and, \ for all \ $d \geq 1$\,,}
  \\[1ex]
    \varphi_{<}^{d+1}(x_d,\twodots,x_0,y_d,\twodots,y_0) 
  & \deff
  & x_d{<}y_d\,\oder\,\big(x_d{=}y_d \und 
    \varphi_{<}^d(x_{d-1},\twodots,x_0,y_{d-1},\twodots,y_0) \big)\,.
  \end{array}
\]}%
\bigskip\\ 
{\bf\em (b):}\quad
We first concentrate on $d{=}1$. 
For ${x_0,y_0,z_1,z_0}\in\dom{N}\deff\set{0,\twodots,N}$, the formula 
$\varphi_{+}^1(x_0,y_0,z_1,z_0)$
shall express that ${x_0}+{y_0}\, = \,{z_1}{\cdot}(N{+}1) + {z_0}$.
\\
If ${x_0}+{y_0}\leq N$, then ${x_0}+{y_0}= {z_1}{\cdot}(N{+}1) + {z_0}$ iff
${z_1}=0$ and ${z_0}={x_0}+{y_0}$.
Otherwise, we have $N{+}1\leq {x_0}+{y_0} < 2{\cdot}(N{+}1)$, and hence there are
${u},{v} \in \dom{N}$ such that 
${x_0}+{u}=N$ and ${u}+{v}={y_0}-1$. Thus 
${x_0}+{y_0} = {x_0}+{u}+1+{v} = (N{+}1) +{v}$. 
Hence, ${x_0}+{y_0}= {z_1}{\cdot}(N{+}1) + {z_0}$ iff
${z_1}=1$ and ${z_0}={v}$.\\ 
This can easily be expressed in $\FO(+)$.
\parno
For $d\geq 1$ the formula $\varphi_+^{d+1}$ can be defined by induction:
Obviously, 
{
\begin{displaymath}
  \begin{array}{c}
    \Nr_N(x_d,x_{d-1},\twodots,x_0) \ +\ 
    \Nr_N(y_d,y_{d-1},\twodots,y_0) \ \, = \ \,
    \Nr_N(z_{d+1},z_d,z_{d-1},\twodots,z_0)
  \medskip\\
    \mbox{if and only if}
  \medskip\\
    \mbox{there is a carry } c\in\set{0,1} \mbox{ such that }
  \\
    \Nr_N(x_{d-1},\twodots,x_0) \ +\ 
    \Nr_N(y_{d-1},\twodots,y_0) \ \, = \ \,
    \Nr_N(c,z_{d-1},\twodots,z_0)
  \\
    \mbox{and } \ c + x_d + y_d \ = \ z_{d+1}{\cdot}(N{+}1) + z_d\,.
  \end{array}
\end{displaymath}}%
The formula $\varphi_+^{d+1}$ can now be easily defined in $\FO(+)$ by
using the formulas $\varphi_+^d$ and $\varphi_+^1$.
\bigskip\\
{\bf\em (c):}\quad
By definition we have for arbitrary numbers $x,y\in\NN$ that
$x\equiv_n y$ iff there is some $z\in\set{0,\twodots,\max(x,y)}$ such that $x= y + n{\cdot}z$
or $y= x + n{\cdot}z$. Since $n$ is fixed, $n{\cdot}z$ can be
expressed by the $n$-fold sum $z+\cdots+z$. When considering
$d$-tuples representing the numbers $x,y$, then $z$ can be represented
by a $d$-tuple, too. By applying \emph{(b)}, the $n$-fold
sum of those $d$-tuples can be expressed in $\FO(+)$. This gives
us the desired formula $\varphi_{\equiv_n}^{d}$.
\bigskip\\
Since the proof of part \emph{(d)} is a bit lengthy, let us first
concentrate on
\bigskip\\
{\bf\em (e):}\quad
For $y\in\dom{N}$ and $\vek{x} \deff (x_{d-1},\twodots,x_0)\in (\dom{N})^d$ the formula
$\varphi_{\Bit}^d(\vek{x},y)$ shall express that the $y$-th bit in the binary 
representation of $\Nr_N(\vek{x})$ is $1$. \\
Because of
$\Nr_N(\vek{x}) \,\deff\, \sum_{i=0}^{d-1}x_i (N{+}1)^i\, < \,(N{+}1)^d$, 
the $y$-th bit of $\Nr_N(\vek{x})$
can be 1 only if $y< d \cdot \lg (N{+}1)$.
For $y < d \cdot \lg (N{+}1)$ let $y = y_1 +\cdots + y_d$ where $y_i<\lg (N{+}1)$,
and let $z_i \deff 2^{y_i} \in\dom{N} $ for $i\in\set{1,\twodots,d}$.
The $y$-th Bit of $\Nr_N(\vek{x})$ is 1 iff $\abgerundet{\frac{\Nr_N(\vek{x})}{2^y}} = 
\abgerundet{\frac{\Nr_N(\vek{x})}{2^{y_1}\cdots 2^{y_d}}} =
\abgerundet{\frac{\Nr_N(\vek{x})}{z_1\cdots z_d}}$ is odd.
\\
Because of $\abgerundet{\frac{a}{bc}} =
\lfloor{\frac{\lfloor{\frac{a}{b}}\rfloor}{c}}\rfloor$ we hence obtain
that the $y$-th Bit of $\Nr_N(\vek{x})$ is 1 iff there are
$y_1,\twodots,y_d,z_1,\twodots,z_d \in\dom{N}$ such that $y= y_1 + \cdots + y_d$,
and $z_i = 2^{y_i}$ for all $i\in\set{1,\twodots,d}$, and there are
$u^i_{d-1},\twodots, u^i_{0} \in\dom{N}$ for all $i\in\set{0,\twodots,d}$ such that
$(u^0_{d-1},\twodots,u^0_0) = (x_{d-1},\twodots,x_0)$, and
$\Nr_N(u^{i+1}_{d-1},\twodots,u^{i+1}_{0}) = 
\abgerundet{\frac{\Nr_N(u^i_{d-1},\twodots,u^i_0)}{z_{i+1}}}$
for all $i\in\set{0,\twodots,d{-}1}$, and 
$\Nr_N(u^d_{d-1},\twodots,u^d_0)$ is odd.
\\
Making use of Theorem~\ref{theorem:arithmetic} and of parts \emph{(a)},
\emph{(b)}, and \emph{(d)} of Theorem~\ref{theorem:d-arithmetic}, this
can easily be expressed by a $\FO(\Bit)$-formula $\varphi^d_{\Bit}$.
\bigskip\\
%
{\bf\em (d):}\quad 
The proof is by induction on $d$. 
In \emph{Step 1} we prove the induction step from $d$ to $d{+}1$, and in \emph{Step 2} we
concentrate the induction start for $d=1$.
\medskip\\
\emph{Step 1:} \ 
Let $d\geq 1$, and assume that the formulas $\varphi^1_{\times}$ and
$\varphi^d_{\times}$ are already available. Our aim is to construct
the formula $\varphi^{d+1}_{\times}$.
\\
To expose the overall idea, we consider the multiplication of two 
decimal numbers. For example, 
\ \ $5731 \times 2293 \ \ = \ \ (5000 + 731)\  \times \ (2000 + 293)\ \ =$
{
\[
     (5000 \times 2000) \ \ 
 + \ \ (5000 \times 293) \ \ 
  + \ \ (731 \times 2000) \ \ 
  + \ \ (731 \times 293)\,.
\]}%
In the same way for numbers \,$\vek{x},\vek{y}\in (\dom{N})^{d+1}$\, it obviously holds that
{
\[
\begin{array}{llcllllr}
  \multicolumn{7}{l}{\Nr_N(x_d,x_{d-1},\twodots,x_0) \ \times\ 
    \Nr_N(y_d,y_{d-1},\twodots,y_0) \ \  = \ \ }
  \\[1ex]
    \qquad\qquad
  & 
  & \big(
  & \Nr_N(x_d,0,\twodots,0)
  & \times
  & \Nr_N(y_d,0,\twodots,0)
  & \big)
  & \qquad\ \mbox{[line 1]}
  \\[0.5ex]
  & + \  
  & \big(
  & \Nr_N(x_d,0,\twodots,0)
  & \times
  & \Nr_N(y_{d-1},\twodots,y_0) 
  & \big)
  & \qquad\ \mbox{[line 2]}
  \\[0.5ex] 
  & +
  & \big(
  & \Nr_N(x_{d-1},\twodots,x_0) 
  & \times
  & \Nr_N(y_d,0,\twodots,0)
  & \big)
  & \qquad\ \mbox{[line 3]}
  \\[0.5ex]
  & +
  & \big(
  & \Nr_N(x_{d-1},\twodots,x_0) 
  & \times
  & \Nr_N(y_{d-1},\twodots,y_0)
  & \big)
  & \qquad\ \mbox{[line 4]}
\end{array}
\]}%
The multiplication in line 1 can be done via the formula
$\varphi^1_{\times}$ as follows:
Let $u^{(1)}_{2d+1}$ and $u^{(1)}_d$ be the numbers which satisfy
\,$\varphi^1_{\times}(x_d,y_d,u^{(1)}_{2d+1},u^{(1)}_d)$, and let
$u^{(1)}_{d-1} = \cdots = u^{(1)}_0 = 0$. Obviously, 
\,$\Nr_N(u^{(1)}_{2d+1},u^{(1)}_{2d},u^{(1)}_{2d-1},\twodots,u^{(1)}_0)$\, is the
result of the multiplication in line 1.
\\
The multiplication in line 2 can be done as follows: 
The formula $\varphi^d_{\times}$ helps to determine numbers
\,$u^{(2)}_{2d+1},\twodots,u^{(2)}_{d}$\, such that
\ $\Nr_N(u^{(2)}_{2d+1},\twodots,u^{(2)}_{d}) \, = \,
 x_d \times \Nr_N(y_{d-1},\twodots,y_0)$. Furthermore, let
$u^{(2)}_{d-1} = \cdots = u^{(2)}_0 = 0$.
Clearly, 
\,$\Nr_N(u^{(2)}_{2d+1},u^{(2)}_{2d},u^{(2)}_{2d-1},\twodots,u^{(2)}_0)$\, is the
result of the multiplication in line 2.
\\
Analogously we obtain a tuple $\vek{u}^{(3)}$ such
that
\,$\Nr_N(u^{(3)}_{2d+1},u^{(3)}_{2d},u^{(3)}_{2d-1},\twodots,u^{(3)}_0)$\,
is the result of the multiplication in line 3.
\\
Furthermore, $\varphi^d_{\times}$ directly gives us a tuple
$\vek{u}^{(4)}$ such that
\,$\Nr_N(u^{(4)}_{2d-1},\twodots,u^{(4)}_0)$\,
is the result of the multiplication in line 4.
\\
For the addition of the numbers
\ $\Nr_N(\vek{u}^{(1)}) + \Nr_N(\vek{u}^{(2)}) + 
   \Nr_N(\vek{u}^{(3)}) + \Nr_N(\vek{u}^{(4)})$ \ 
we make use of part \emph{(b)} of Theorem~\ref{theorem:d-arithmetic}.
\\
Altogether, this gives us the desired $\FO(+,\times)$-formula
$\varphi^{d+1}_{\times}$.
\bigskip
\\
\emph{Step 2:} \ 
We now construct the $\FO(+,\times)$-formula 
$\varphi^{1}_{\times}(x,y,z_1,z_0)$, expressing that
\ $x\cdot y \ = \ z_1 \cdot (N{+}1) + z_0$.
\\
Let \,$M\deff\lfloor\sqrt{N}\rfloor$.
The basic idea is the following:
\begin{enumerate}
\item 
  Move over from numbers \,$x\in\set{0,\twodots,N}$\, to their
  $(M{+}1)$-ary representations \,$x = u_1\cdot (M{+}1) + u_0$\, for 
  \,$u_1,u_0 \in \set{0,\twodots,M}$.
  Note that this is possible for all $x\leq N$ since
  \,$M^2 \leq N < (M{+}1)^2 = M\cdot (M{+}1) + M +1$.
\item 
  Show that this decomposition can be defined by a $\FO(+,\times)$-formula
  $\chi(x,u_1,u_0)$ which expresses that 
  \ $x = \Nr_M(u_1,u_0)$.
\item
  Construct a formula
  \,{$\psi^2_{\times}(u_1,u_0,v_1,v_0,w_3,w_2,w_1,w_0)$}\, that defines the 
  multiplication for the $(M{+}1)$-ary representations. \\ I.e., $\psi^2_{\times}$
  expresses that
  \ $\Nr_M(u_1,u_0)\times \Nr_M(v_1,v_0) \  = \ 
    \Nr_M(w_3,w_2,w_1,w_0)$.
\item
  Show that $\chi$ can be extended to a formula 
  $\chi'(z_1,z_0,w_3,w_2,w_1,w_0)$ which expresses that 
  $\Nr_N(z_1,z_0)\  = \ \Nr_M(w_3,w_2,w_1,w_0)$.
\end{enumerate}
Afterwards, the desired formula $\varphi^1_{\times}$ can be defined as follows:
{\setlength{\arraycolsep}{0ex}\small\[
  \begin{array}{rcll}
    \varphi^1_{\times}(x,y,z_1,z_0)
  & \ \ \deff\ \  
  & \exists u_1, u_0, v_1, v_0 , w_3, w_2, w_1, w_0 \ \ \big(\ 
  & \chi(x,u_1,u_0)\ \und\ \chi(y,v_1,v_0)\ \und
  \\[0.5ex]
  & & & \psi^2_{\times}(u_1,u_0,v_1,v_0,w_3,w_2,w_1,w_0)
  \\[0.5ex]
  & & & \chi'(z_1,z_0,w_3,w_2,w_1,w_0)\ \big)\,.
  \end{array}
\]}%
Hence it suffices to construct the formulas $\chi$, $\psi^2_{\times}$, and $\chi''$.
\\
Of course, the formula
\begin{eqnarray*}
  \zeta_M(z) & \deff &
  \exists y\; \big(\, z\times z = y \ \und\ 
    \forall z'\; (\,
        z'>z\ \impl\ \nicht\exists y'\, z'\times z'=y'
      \,)\,\big)
\end{eqnarray*}
expresses, for underlying universe
\,$\set{0,\twodots,N}$, that the variable $z$ is interpreted with the number 
\,$M\deff\lfloor\sqrt{N}\rfloor$.
Consequently, the following formula \,$\chi(x,u_1,u_0)$\, expresses
that \,$x= \Nr_M(u_1,u_0) = \ u_1\cdot (M{+}1) + u_0$\,:
{\setlength{\arraycolsep}{0ex}
\[
\begin{array}{rcll}
  \chi(x,u_1,u_0) 
& \ \,\deff\,\  
& \exists z \,\exists v\,\exists w\ \;\big(\ 
& \zeta_M(z) \ \und \ u_1\leq z \ \und\ u_0\leq z\ \und\ 
\\
& & & v=u_1\times z \ \und\ w = v + u_1\ \und\ x = w + u_0\ \big)\,.  
\end{array}
\]}%
It is straightforward to obtain a formula \,$\psi^1_{\times}(u,v,w_1,w_0)$\, which expresses,
for underlying universe \,$\set{0,\twodots,N}$\, and interpretations of \,$u,v,w_1,w_0$\,
in \,$\set{1,\twodots,M}$, \,that \,$u\cdot v = w_1\cdot (M{+}1) + w_0$\,:
\begin{eqnarray*}
 \psi^1_{\times}(u,v,w_1,w_0) & \deff &
 \exists w\ \;\big(\ \chi(w,w_1,w_0)\ \und\ w= u\times v\ \big)\,.
\end{eqnarray*}
In the same way as in \emph{Step 1} we obtain a formula
\,$\psi^2_{\times}(u_{1},u_0,v_{1},v_0,w_3,w_{2},w_1,w_0)$\,
which expresses, 
for underlying universe \,$\set{0,\twodots,N}$\, and interpretations
of \,$\vek{u},\vek{v},\vek{w}$\,
in \,$\set{1,\twodots,M}$, \,that
\[
    \Nr_M(u_1,u_0) \ \times\ 
    \Nr_M(v_1,v_0) \ \, = \ \,
    \Nr_M(w_3,w_2,w_1,w_0)\,.
\]
All that remains to do is to construct a formula 
\,$\chi'(z_1,z_0,w_3,w_2,w_1,w_0)$\, which expresses that 
$\Nr_N(z_1,z_0)\ = \ z_1\cdot (N{+}1) + z_0\ = \ \Nr_M(w_3,w_2,w_1,w_0)$.
\\
Choose \,$a_1,a_0,b_1,b_0,n_1,n_0$\, such that
\,$z_1 = \Nr_M(a_1,a_0)$, 
\ $z_0 = \Nr_M(b_1,b_0)$, \,and
\ $N{+}1= \Nr_M(n_1,n_0)$.
For $z_1$ and $z_0$ this can be ensured by the formula $\chi$; for $N{+}1$ it can
be ensured by a straightforward variant of $\chi$.\\
Obviously, 
\ {\small $z_1{\cdot} (N{+}1) + z_0 
   \,=\, 
   \big(\, \Nr_M(a_1,a_0) \times \Nr_M(n_1,n_0) \,\big) + \,
   \Nr_M(b_1,b_0)$}\,. 
Hence, 
\[
  \begin{array}{c}
      z_1\cdot (N{+}1) + z_0\ \ = \ \ \Nr_M(w_3,w_2,w_1,w_0)
   \\[1ex]
     \mbox{if and only if}
   \\[1ex]
     \big(\ \Nr_M(a_1,a_0)\ \times\ \Nr_M(n_1,n_0)\ \big)\ +\ \Nr_M(b_1,b_0)
     \ \ = \ \ \Nr_M(w_3,w_2,w_1,w_0)\,.
  \end{array}
\]
This multiplication and addition of $(M{+}1)$-ary representations can be done in 
a straightforward way by using the formula 
$\psi^2_{\times}$ and an according formula $\psi^4_{+}$ (obtained by a 
variant of part \emph{(b)} of Theorem~\ref{theorem:d-arithmetic}).
Altogether, this gives us the desired formula $\chi'$.
\\
%
%
Finally, the proof of part \emph{(d)} and, altogether, the proof of 
Theorem~\ref{theorem:d-arithmetic} is complete.
\end{proofc_mit}%
%


\end{document}